\documentclass{aa} 
\usepackage{graphicx}

\newcommand{\orion} {{\sc M42}}
\newcommand{\Teff} {{\em T}$_{\rm eff}$}
\newcommand{\grav} {log\,{\em g}}
\newcommand{\vsini} {$v$\,sin\,$i$}
\newcommand{\fastwind} {{\sc Fastwind}\ }
\newcommand{\iraf} {{\sc iraf}\ }
\newcommand{\fwhm} {{\sc fwhm}\ }
\newcommand{\kms} {km\,s$^{-1}$}
\newcommand{\HII} {{\sc H\,ii}\ }

\newcommand{\feros} {{\sc feros}}
\newcommand{\newton} {{\sc int}}
\newcommand{\ids} {{\sc ids}}
\newcommand{\idl} {{\sc idl}}
\newcommand{\snr} {{\em SNR}}
\newcommand{\ew} {{\em EW}}
\newcommand{\radius} {{\em R}}
\newcommand{\mass} {{\em M}}
\newcommand{\lum} {{\em L}}
\newcommand{\vinf} {$v_\infty$}

\begin{document}


\title{Detailed spectroscopic analysis of the Trapezium cluster stars 
       inside the Orion Nebula}
\subtitle{Rotational velocities, stellar parameters and oxygen abundances
          \thanks{The \newton\ is operated on the island of La Palma by 
		  the RGO in the Spanish Observatorium of El Roque de los 
		  Muchachos of the Instituto de Astrof\'{\i}sica de Canarias.}}

\author{S. Sim\'on-D\'{\i}az \inst{1,2}, A. Herrero\inst{1,3}, 
        C. Esteban \inst{1,3} and F. Najarro\inst{4}}

\institute{Instituto de Astrof\'\i sica de Canarias, E-38200 La Laguna, 
           Tenerife, Spain
		   \and
		   Isaac Newton Group of Telescopes, Apartado de Correos 321, 
		   E-38700 Santa Cruz de la Palma, Tenerife, Spain.
           \and
           Departamento de Astrof\'{\i}sica, Universidad de La Laguna, 
	       Avda. Astrof\'{\i}sico Francisco S\'anchez, s/n, E-38071 
	       La Laguna, Spain
		   \and
		   Instituto de Estructura de la Materia, CSIC, C/ Serrano 121, 
		   28006 Madrid, Spain} 

\offprints{ssimon@iac.es}

\date{Submitted/Accepted}
\titlerunning{Spectroscopic analysis of the Trapezium cluster 
              stars}
\authorrunning{S. Sim\'on-D\'{\i}az et al.}

\abstract{We present the results of a spectroscopic analysis of the
Trapezium cluster stars inside the Orion Nebula. The rotational velocities
have been obtained using Fourier analysis method, finding agreement with
values derived from the usual method, based on linewidth measurements.
The rotational velocity derived for $\theta^1$ Ori C through this method
is consistent with the variability of some of its spectral features that 
have a period of 15.42 days.
By means of the fit of \ion{H}{}, \ion{He}{i} and \ion{He}{ii} observed
profiles with \fastwind synthetic profiles, stellar parameters and wind
characteristics have been derived. This methodology let us estimate the 
errors associated with these parameters.
It is found that macroturbulence effects have to be included for a good
fit to the \ion{He}{i-ii} lines in the spectrum of $\theta^1$ Ori C.\\
By means of a very accurate study, oxygen abundances have been derived 
for the three B0.5V  stars $\theta^1$ Ori A, D and 
$\theta^2$ Ori B. Final abundances are consistent with the nebular gas-phase 
results presented in Esteban et al. (\cite{Est04}) and are lower than
those given by Cunha \& Lambert (\cite{Cun94}). Our results suggest
a lower dust depletion factor of oxygen than previous estimations for the
Orion nebula.\\

}
\maketitle

\section{Introduction}
New developments of massive star model atmosphere codes have led to 
interesting new possibilities for stellar spectroscopic studies.
Improvements in computational methods as well as an increase of the 
efficiency of computers, have made possible the modeling of atmospheres
of hot luminous stars, taking into account not only strong NLTE effects and 
hundreds thousands metallic lines producing the so-called {\em line-blanketing}
(Hubeny \& Lanz \cite{Hub95}), but also winds with expanding 
spherical geometries (Santolaya-Rey et al. \cite{San97}; Hillier \& Miller 
\cite{Hil98}; Pauldrach et al. \cite{Pau01}; Puls et al. \cite{Pul05}).\\
The new improvements that have been included in these latest generation models call
for a review of previous results. For example, the papers by Herrero et al. 
(\cite{Her02}), Crowther et al. (\cite{Cro02}), and Martins et al. (\cite{Mar02}) showed 
that the SpT - \Teff\ calibrations used previously (Vacca et al. \cite{Vac96})
needed to be revised to lower effective 
temperatures for a given spectral type. Recent analyses by Repolust et al. 
(\cite{Rep04}) and Martins et al. (\cite{Mar05}) in the Milky Way, and by 
Massey et al. (\cite{Mas04}, \cite{Mas05}) at SMC and LMC 
metalicities, reinforce this 
result and together implies a need to revisit the ionizing flux  distribution 
that is used for the study of \ion{H}{ii} regions and starbursts.\\
\newline
Recent work by Trundle et al. (\cite{Tru02}) and Urbaneja et al. (\cite{Urb05}) 
have shown that abundance gradients in some spiral galaxies derived from stellar 
and nebular studies tend to be coherent but very dependent on the calibration 
used in the strong line nebular methods. However, until now, there have 
been no detailed systematic studies which compare results from nebular 
and stellar studies. This is the first of a series of papers aimed at this 
subject. We have selected some 
galactic \ion{H}{ii} regions for a detailed study of the interaction 
between massive stars and the surrounding ISM, looking for consistency of
derived parameters (\Teff, luminosities and ionising flux distribution of 
the stellar population) as well as abundances of C, N, O, Si and Mg. 
For this first study we have selected the Orion nebula, a well studied and 
resolved Galactic \ion{H}{ii} region powered by a cluster of a few massive stars, 
the Trapezium cluster.\\
\newline 
The Orion complex contains the massive on-going star forming region closest 
to Earth, at about only 450 pc. The Orion nebula, \orion, is part of this 
complex. It is a well known \HII region (e.g. O'\,Dell \cite{Ode01}; Ferland 
\cite{Fer01}) ionised by the Trapezium cluster stars ($\theta^1$ Ori), a 
group of early type stars located in the core of the nebula. Together with 
$\theta^1$\,Ori\,C (HD\,37022, O7V), the main ionising source, we find other 
B0.5V stars, perfect targets for a stellar abundance study.\\
The most recent study of the chemical composition of the Orion nebula has 
been presented by Esteban et al. (\cite{Est04}) who reanalysed the emission 
line spectrum of the nebula to determine the physical conditions and
abundances of the ionised gas-phase. Cunha \& Lambert (\cite{Cun92}, 
\cite{Cun94}) included some of the Trapezium cluster stars in a survey of 
B-type stars in the Orion OB1 association. They presented a spectroscopic 
analysis of these stars for determining C, N, O, Si and Fe stellar abundances.\\
\newline
This paper is focused on the 
spectroscopic analysis of the Trapezium cluster stars for deriving their 
stellar parameters and oxygen abundances. The stellar 
parameters obtained for $\theta^1$\,Ori\,C will be used in future papers 
as input for the 
modeling of the Orion Nebula with photoionization codes. The derived 
stellar abundances are compared to those obtained by Esteban et al. 
(\cite{Est04}) through nebular studies.\\
\newline
Our paper is structured as follows: In Sect. \ref{observations} we present 
the observations. In Sect. \ref{rotational} and \ref{parameters} we obtain 
the \vsini\ and the stellar parameters of our targets. The study of 
$\theta^1$ Ori C is presented in Sect. \ref{HD37022}, 
and the oxygen abundance analyses in Sect. \ref{abundance}. 
A discussion of the results and the conclusions of this work are
presented in Sect. \ref{conclude}.
\begin{figure*}[ht!]
\centering
\includegraphics[width=13. cm,angle=90]{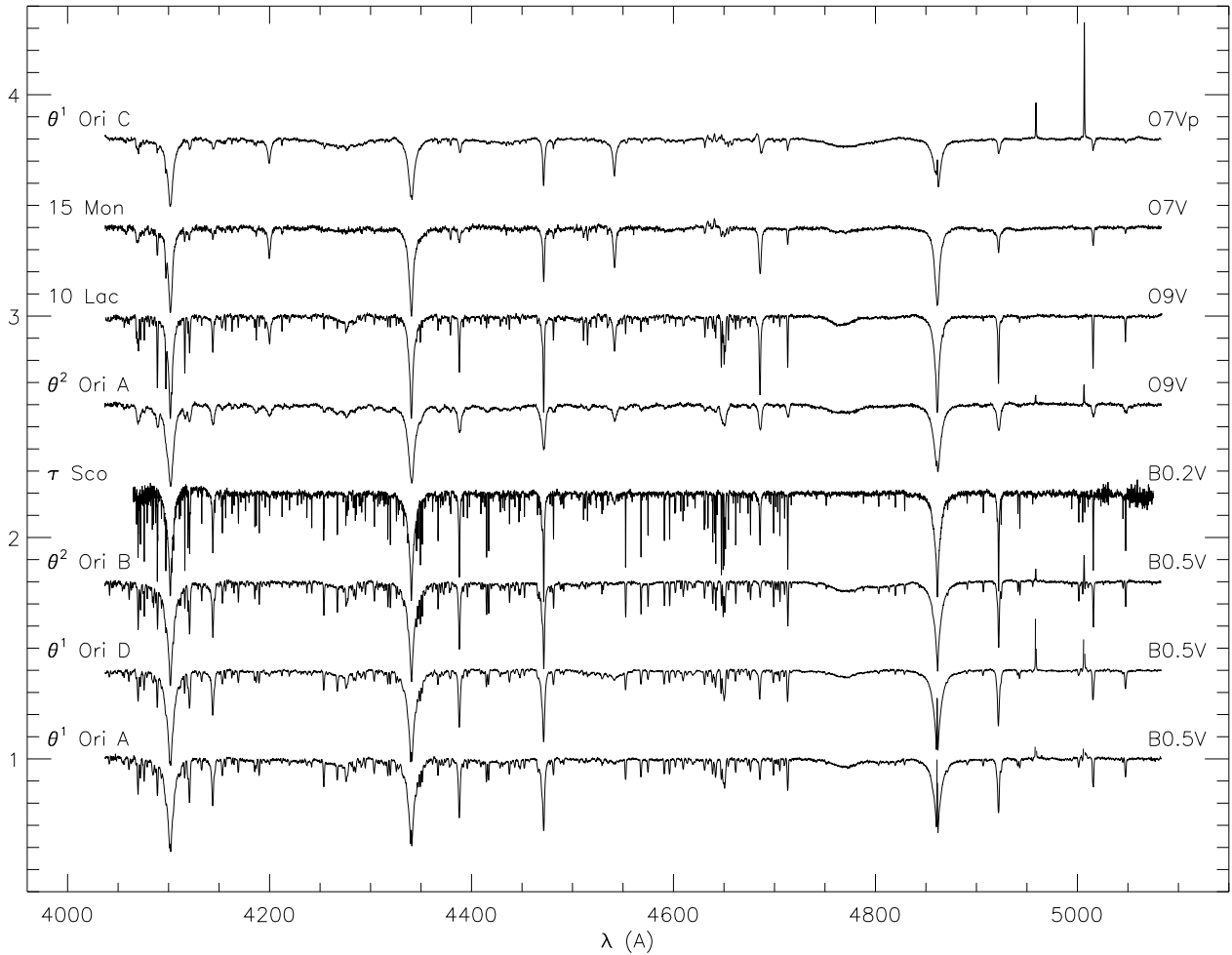}
\caption{Atlas of the \newton+\ids\ spectra in the 4000-5000 {\rm \AA}\ region 
         used for this study. $\tau$ Sco spectrum is from {\sc caspec}. Several 
		 narrow nebular emission lines appear in the spectra of the Orion stars. 
		 In these stars, the cores of the hydrogen Balmer lines are contaminated 
		 by the nebular lines. Nebular H$_{\rm \beta}$ and  
         [\ion{O}{iii}]\,$\lambda\lambda$\,4960,5007 lines have been arbitrarily 
		 diminished in the $\theta^1$ Ori A and D plotted spectra for clarity 
		 reasons. Many metal lines can be distinguished together with the 
		 broad \ion{H}{} and \ion{He}{} lines in the O9 - B0.5 spectra. The high 
		 rotational velocity of $\theta^2$ Ori A makes the metal lines of this 
		 O9V star shallower, broader and blended. The known inverted P-Cygni
		 \ion{He}{ii}\,$\lambda$4686 profile can be seen in the $\theta^1$ 
		 Ori C spectrum.}\label{espectros4000-5000}
\end{figure*}
\begin{figure}[ht!]
\centering
\includegraphics[width=14. cm,angle=90]{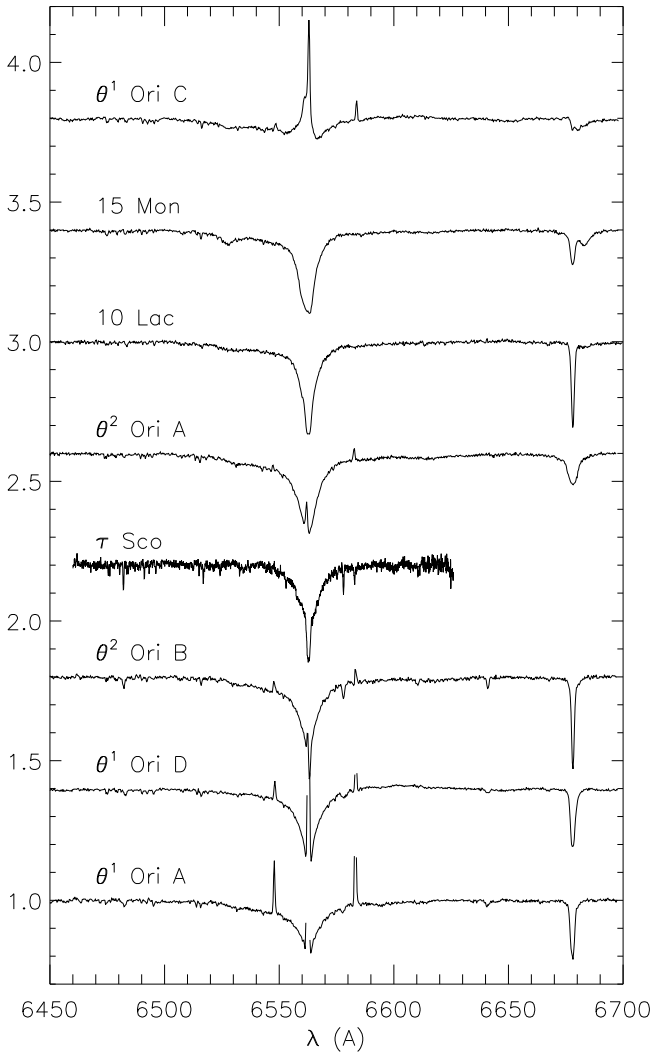}
\caption{Atlas of the \newton+\ids\ spectra in the H$_\alpha$ region used for this 
         study. $\tau$ Sco spectrum is from {\sc caspec}. Nebular H$_\alpha$ 
		 and [\ion{N}{ii}] emission lines have been diminished in the 
		 $\theta^1$ Ori A and D plotted spectra for clarity reasons. Two 
		 emission components can be distinguished in the $\theta^1$ 
		 Ori C H$_\alpha$ line: the narrow one is associated with the nebula, 
		 the other one is variable and it is associated with the star (see 
		 section \ref{HD37022}). The blue wing of the H$_{\alpha}$ line
		 is affected by some cosmetic in the \newton\ observations.}
		 \label{espectrosHa}
\end{figure}
\section{The observations}
\label{observations}
The bulk of the observations used here were carried out with the Isaac 
Newton 2.5m. Telescope (\newton) at the Roque de los Muchachos observatory in La 
Palma on the 21th of December 2002. The Intermediate Dispersion Spectrograph 
(\ids) was used with the 235 mm camera and two different gratings. We have observed 
the spectral region between 4000 and 5050 \AA \ using the H2400B grating, 
which resulted in an effective spectral resolution R\,$\sim$\,7500 
(equivalent to a 0.23 \AA/pixel resolution and $\sim$ 2.6 pixel \fwhm 
arc lines). For the H$_{\alpha}$ region the H1800V grating was used, 
resulting in a similar spectral resolution (0.3 \AA/pixel, R\,$\sim$\,8000). 
With the configurations used three exposures were needed to cover the whole 
range. A large number of flat fields and arcs for the reduction process were
obtained.\\
\newline
The reduction and normalization of the spectra was made following standard 
techniques, with \iraf and our own software developed in \idl. The 
signal-to-noise ratio (\snr) of the reduced spectra depends on the 
spectral range, but is usually about 200-250 in the blue region and 
250 in the H$_{\alpha}$ region (see Table \ref{SNR}).\\
\newline
We found some problem when rectifying the \newton\ observations
in the H$_{\alpha}$ region. In the blue wing of the H$_{\alpha}$ 
line we observed a feature which is independent of rectification. 
No known line should be present at this wavelength, so we argue 
that it must be cosmetic.
\\
\newline
Special care has to be taken over the nebular contamination of the stellar 
spectra. The stars studied are located inside \ion{H}{ii} regions, so the 
stellar spectra are contaminated by the nebular emission spectrum. 
It would be desirable to remove this nebular contribution, as it fills 
the cores of the Balmer \ion{H}{} and \ion{He}{i} lines. However this is 
not easy even though we have long slit 
observations; the nebular emission has spatial structure that complicates 
its subtraction (especially for the Balmer \ion{H}{} lines, the most 
important nebular contribution). If this is not correctly done an over or under 
subtraction will appear. After trying different possibilities we concluded that 
the best solution for this problem is not to subtract the emission 
lines and to ignore these regions in the final spectrum. For the Balmer \ion{H}{} 
and \ion{He}{i} lines this is satisfactory, as emission lines are narrower than 
absorption lines. Nebular contributions could be more difficult to separate 
for metal lines, however the contamination of the stellar metal lines used for
the abundance analysis due to nebular lines is negligible.\\
\begin{table}[h!]
\begin{center}
\scriptsize{
\begin{tabular}{c c c c c c}    
\noalign{\smallskip}
\hline
\noalign{\smallskip}
 HD\, & Name & SpT & m$_{\rm v}$ & A$_{\rm v}$ & M$_{\rm v}$ \\
\hline
\noalign{\smallskip}
Orion stars & & & & & \\
\hline
\noalign{\smallskip}
 HD\,37020  & $\theta^1$ Ori A & B0.5V  & 6.73 & 1.89 & -3.4 \\
 HD\,37022  & $\theta^1$ Ori C &  O7Vp  & 5.12 & 1.74 & -4.9 \\
 HD\,37023  & $\theta^1$ Ori D &  B0.5V & 6.71 & 1.79 & -3.3 \\
 HD\,37041  & $\theta^2$ Ori A &  O9V   & 5.07 & 1.12 & -4.3 \\
 HD\,37042  & $\theta^2$ Ori B &  B0.5V & 6.41 & 0.73 & -2.6 \\
\noalign{\smallskip}
\hline
\noalign{\smallskip}
Reference stars & & & & & \\
\hline
\noalign{\smallskip}
 HD\,47839  & 15 Mon           &  O7V   &  --- & ---  & -4.8 \\
 HD\,214680 & 10 Lac           &  O9V   &  --- & ---  & -4.4 \\
 HD\,149438 & $\tau$ Sco       &  B0.2V &  --- & ---  & -3.3 \\
\noalign{\smallskip}
\hline
\\
\end{tabular}
}
\normalsize
\rm
\caption{\footnotesize Identification, spectral type and photometric 
visual data of the studied objects. $A_{\rm v}$ and $m_{\rm v}$ values for Orion 
stars from Preibisch et al. (\cite {Pre99}). $M_{\rm v}$ values for these 
stars have been calculated considering a distance $d\,\sim$\,450\,$\pm$\,50 pc 
to the Orion nebula. Data for HD\,214680 and HD\,47839 are from
Herrero et al. (\cite{Her92}). Photometric data for HD\,149438 are from Humphreys
(\cite{Hum78}). Uncertainties in $m_{\rm v}$, $A_{\rm v}$ and $M_{\rm v}$ are
0.01, 0.03 and 0.3 respectively.}\label{t1}
\end{center}
\end{table}
\begin{table}[h!]
\begin{center}
\scriptsize{
\begin{tabular}{c c c c}    
\noalign{\smallskip}
\hline
\noalign{\smallskip}
 HD\, & 4000 - 5000 & 4600 - 5100 & H$_{\rm \alpha}$ \\
\noalign{\smallskip}
\hline
\noalign{\smallskip}
 HD\,37020  & 170-280 & 210 & 220 \\
 HD\,37022  & 170-330 & 280 & 290 \\
 HD\,37023  & 220-450 & 450 & 320 \\
 HD\,37041  & 200-275 & 230 & 210 \\
 HD\,37042  & 160-220 & 250 & 230 \\
 HD\,214680 & 140-190 & 200 & 160 \\
 HD\,47839  & 89-120  & 160 & 260 \\
\noalign{\smallskip}
\hline
\\
\end{tabular}
}
\normalsize
\rm
\caption{\footnotesize \snr\ achieved for the different 
         spectra for the three ranges observed with the \newton+\ids.}
		 \label{SNR}
\end{center}
\end{table}
\newline
The \newton\ observations consist of the brightest three stars 
in the Trapezium cluster ($\theta^1$ Ori A, C, D) together with the two 
nearby stars $\theta^2$ Ori A and B. Two standard stars were included 
in this set, 10 Lac and 15 Mon (O9V and O7V respectively).
The other standard star, $\tau$ Sco, was kindly provided by Dr. Gehren. 
This is a slow rotating B0.2V star perfect for a preliminary abundance 
analysis study. The spectrum was obtained with {\sc caspec}, attached 
to the {\sc eso} 3.6m telescope. The \snr\ of this spectrum is 
$\sim$\,200-300 in the blue region and $\sim$\,150 in the 
H$_\alpha$ region.\\
\newline
For the study of the spectral variability of $\theta^1$\,Ori\,C, we have 
used \feros\ spectra (some of them downloaded from the \feros\ database and
other kindly provided by Dr. Stahl). These observations 
were carried out with \feros\ at the {\sc eso} 1.52m telescope in La 
Silla. The instrument is designed for high-dispersion spectroscopy with 
$R\,\sim$\,48000 in the spectral range 3700 - 9200 \AA. The achieved 
\snr\ is 300 at about 4500 - 5000 \AA. Different phase observations have 
been used for the variability study (see Table \ref{tvariable}). This 
study is presented in section \ref{HD37022}.
%
\section{Determination of rotational velocities}
\label{rotational}
The analysis of stellar spectra makes use of a number of free parameters 
like the micro and macroturbulent velocities and the projected rotational 
velocity, \vsini. The last one has acquired particular importance in 
recent times because of the mixing that rotation may induce in the interior 
of massive stars (e.g. Maeder \& Meynet, \cite{Mae00}; Villamariz et al., 
\cite{Vil02}). 
However, some methods to determine the rotational velocities do not distinguish
between rotation and other surface broadening mechanisms, like macroturbulence.\\
\newline
Conventionally, \vsini\ values are based on linewidth measurements of 
individual features, mainly metal lines apparently free of blends. As the 
principal broadening mechanism of these lines is stellar rotation, with 
sufficient resolution it is possible to determine \vsini\ from the \fwhm 
of the line. Usually metal lines are used, however in cases of high 
rotational velocities or high temperatures metal lines appear 
blended or are very week. Therefore, in these cases, if the \vsini\ is 
high, the whole \ion{He}{} spectrum is used; if \vsini\ is not extremely high, 
only \ion{He}{i} lines are used, as these lines are less affected by 
pressure broadening than \ion{He}{ii} lines. However, the \vsini\ 
derived must be tested with some metal lines (if available), as we 
are not completely sure of rotation broadening dominating over pressure 
broadening.\\
\newline
The Fourier method of determining \vsini\ is based on the fact that in the 
Fourier space convolutions transform into products and of the rotation, 
macroturbulence, natural and instrumental profiles (turbulence and 
instrumental assumed gaussian), only the rotation function has zeroes in 
its Fourier transform. These zeroes will appear in the total transform 
function, and Carroll (\cite{Car33}) showed that the position of the zeroes 
are related to the \vsini. Actually the frequency of the first zero 
($\sigma_{1}$) is related to the rotational velocity through:
\begin{equation}
\frac{\lambda}{c}\, \vsini\ \sigma_{1} = 0.660
\end{equation}
The microturbulence also introduces zeros in the Fourier transform at
high frequencies (Gray, \cite{Gra73}). This has to be taken into account 
for very low \vsini\ ($\leq$ 20 \kms).\\
\newline
The main problems in the application of the Fourier method for early type 
stars are related to the quality of the observed spectra (i.e. spectral
resolution and \snr). The lowest \vsini\ limit that can be determined 
is given by the spectral resolution ($\Delta\lambda$ in \AA/pixel), as the 
sampling of the computational Fourier transform cannot be extended beyond
0.5/$\Delta\lambda$.
The noise in the observed spectra transforms as white noise that obscure
the first zero.\\
\newline
The advantages of the Fourier method are that rotational broadening can 
be separated from other broadening mechanisms, and therefore metal lines 
as well as \ion{He}{} lines can be used for the \vsini\ determination 
(even for low values of \vsini). This is very useful for fast rotating 
stars and spectral types earlier than O9, showing blended or very weak 
metal lines.\\
\begin{figure}[t!]
\centering
\includegraphics[width=6.2cm,angle=90]{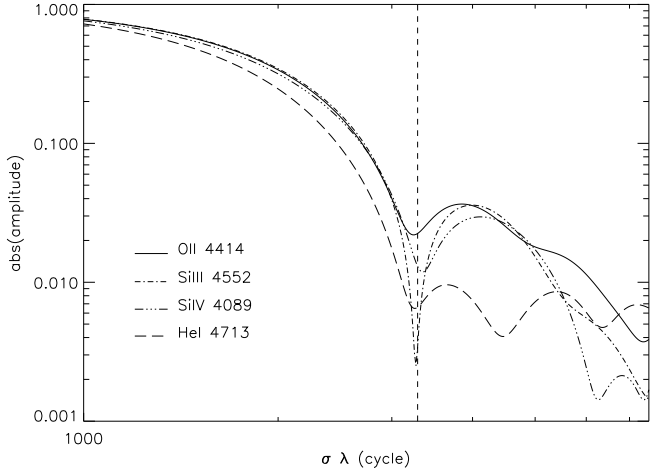}
\caption{Four different synthetic lines generated with \fastwind have been 
         convolved with a \vsini\ of 60 \kms\ and degraded to a \snr\ 
		 of 200. The different lines show the Fourier transform of the 
		 lines on a $\sigma\lambda$ baseline. The first zero is very close 
		 to the theoretical value for all the lines (even for the case of
		 the \ion{He}{i} line.}
\label{fourier_ex}
\end{figure}
\begin{table}[t!]
\begin{center}
\scriptsize{
\begin{tabular}{c c c c c}    
\noalign{\smallskip}
\hline
\noalign{\smallskip}
 HD\, & \multicolumn{3}{c}{\vsini\ (\kms)}  \\
\noalign{\smallskip}
\hline
\noalign{\smallskip}
 & \multicolumn{2}{c}{This work} & Other authors \\
\noalign{\smallskip}
\cline{2-4}
\noalign{\smallskip}
 & Fourier & \fwhm & \fwhm \\
\noalign{\smallskip}
\hline
\noalign{\smallskip}
 HD\,37020  & 55.0\,$\pm$\,0.6 & 57\,$\pm$\,3 & 56$^a$ \\
 HD\,37022  & 24\,$\pm$\,3     & 40\,$\pm$\,11 & 53$^b$ \\
 HD\,37023  & 49.0\,$\pm$\,0.9 & 51\,$\pm$\,3 & 51$^a$ \\
 HD\,37041  & 131\,$\pm$\,4    & 140\,$\pm$\,11 & 110$^b$ \\
 HD\,37042  & 34.0\,$\pm$\,0.5 & 36\,$\pm$\,3 & 10$^c$ \\
 HD\,214680 & 30.0\,$\pm$\,0.8 & 37\,$\pm$\,4 & 35$^b$ \\
 HD\,47839  &  67\,$\pm$\,4    & 66\,$\pm$\,6 & 67$^b$ \\
 HD\,149438 & $<$ 13           & 16\,$\pm$\,3 & 5$^d$, 19$^e$ \\
\noalign{\smallskip}
\hline
\\
\end{tabular}
}
\normalsize 
\rm
\caption{\footnotesize Projected rotational velocity derived from 
         the Fourier and \fwhm analyses. References from the 
		 literature are: $^a$ Sim\'on-D\'{\i}az et al. \cite{Sim03}, 
		 $^b$ Howarth et al. \cite{How97}, $^c$ McNamara \& Larsson
		 \cite{McN62}, $^d$ Sch\"onberner et al. \cite{Sch88}, 
		 $^e$ Killian et al. \cite{Kil91}}\label{t2}
\end{center}
\end{table}
\newline
We have tested the Fourier method with theoretical and observational 
cases, and it works well for massive hot stars. A paper with these 
results as well as determinations for a number of O star rotational 
velocities is in preparation (Sim\'on-D\'{\i}az $\&$ Herrero, 
\cite{Sim05}), figure \ref{fourier_ex} shows a typical example. For 
a recent application to A-stars see Royer et al. (\cite{Roy02}).\\  
\newline
The \vsini\ of our sample of stars has been determined through the 
Fourier and \fwhm methods. Results are presented in Table \ref{t2}, 
together with some values found in the literature, for comparison. 
The resolution in the \ids\ spectra is 0.23 \AA/pixel, so the lowest 
\vsini\ that could be detected through Fourier method is $\sim$20 
\kms. For $\theta^1$ Ori C \feros\ spectra the resolution is 
$\sim$0.03 \AA/pixel, so \vsini\ $\geq$ 2 \kms\ for detection. For 
the $\tau$ Sco spectrum, the resolution is 0.1 \AA/pixel, so the 
lowest detectable \vsini\ is 8 \kms.\\
\newline
All the papers referenced in Table \ref{t2} use the \fwhm method 
applied to the optical spectra of the stars except those by Howarth 
et al. (\cite{How97}), who use a cross correlation technique for 
{\sc iue} spectra, and Sch\"onberner et al. (\cite{Sch88}) who 
compare the observed spectrum of $\tau$ Sco with synthetic profiles.\\  
Comments on the individual stars' \vsini\ determination, as well as the 
comparison between the values derived through Fourier and \fwhm methods 
are presented in Section \ref{individual}. Agreement between both 
methodologies is very good, however there are some interesting cases
(see the study of $\theta^1$ Ori C in Section \ref{HD37022}).
\section{Stellar parameters}\label{parameters}
\begin{figure*}[ht!]
\centering
\includegraphics[width=12.cm,angle=90]{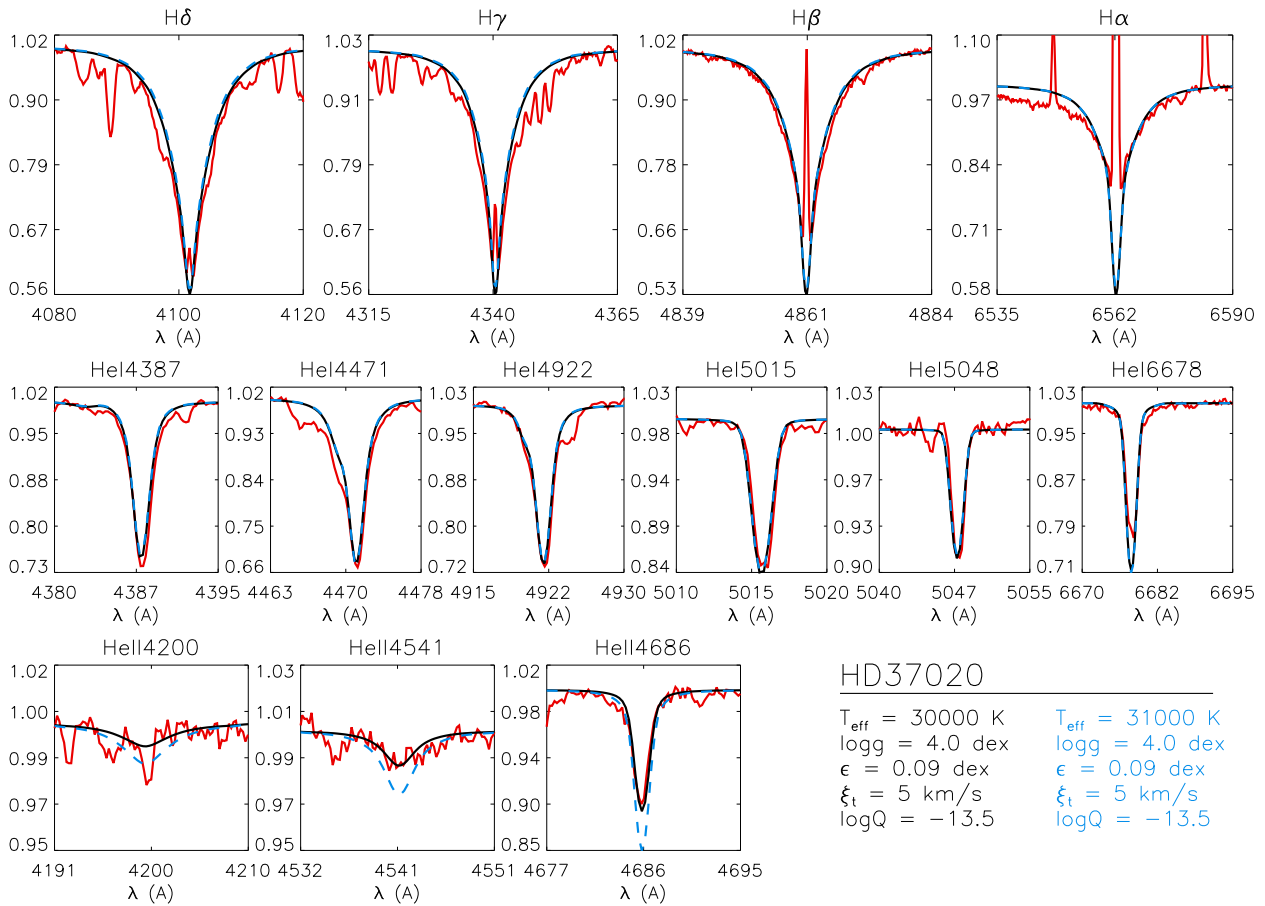}
\caption{Analysis of HD\,37020 ($\theta^1$ Ori A, B0.5V).
		 Two models have been plotted for comparison with the observed 
		 spectrum. The adopted model has been represented with a solid line, 
		 while another model varying 1000 K in \Teff\ has been plotted
         with a dashed line. Note the core of the stellar Balmer lines 
         contaminated with emission from the nebula. Only the wings of 
		 the Balmer lines are used for the fitting. The narrow line that 
		 appears in the core of the \ion{He}{ii}\,4200 line is a \ion{N}{iii} 
         absorption line; the red wing of the \ion{He}{i}\,4922 line is
		 contaminated by an \ion{O}{ii} absorption line. The blue
         wing of the H$_{\alpha}$ line is affected by a cosmetic
		 feature and is not used for the fitting}\label{HD37020_fit}
\end{figure*}

\begin{figure*}[ht!]
\centering
\includegraphics[width=12.cm,angle=90]{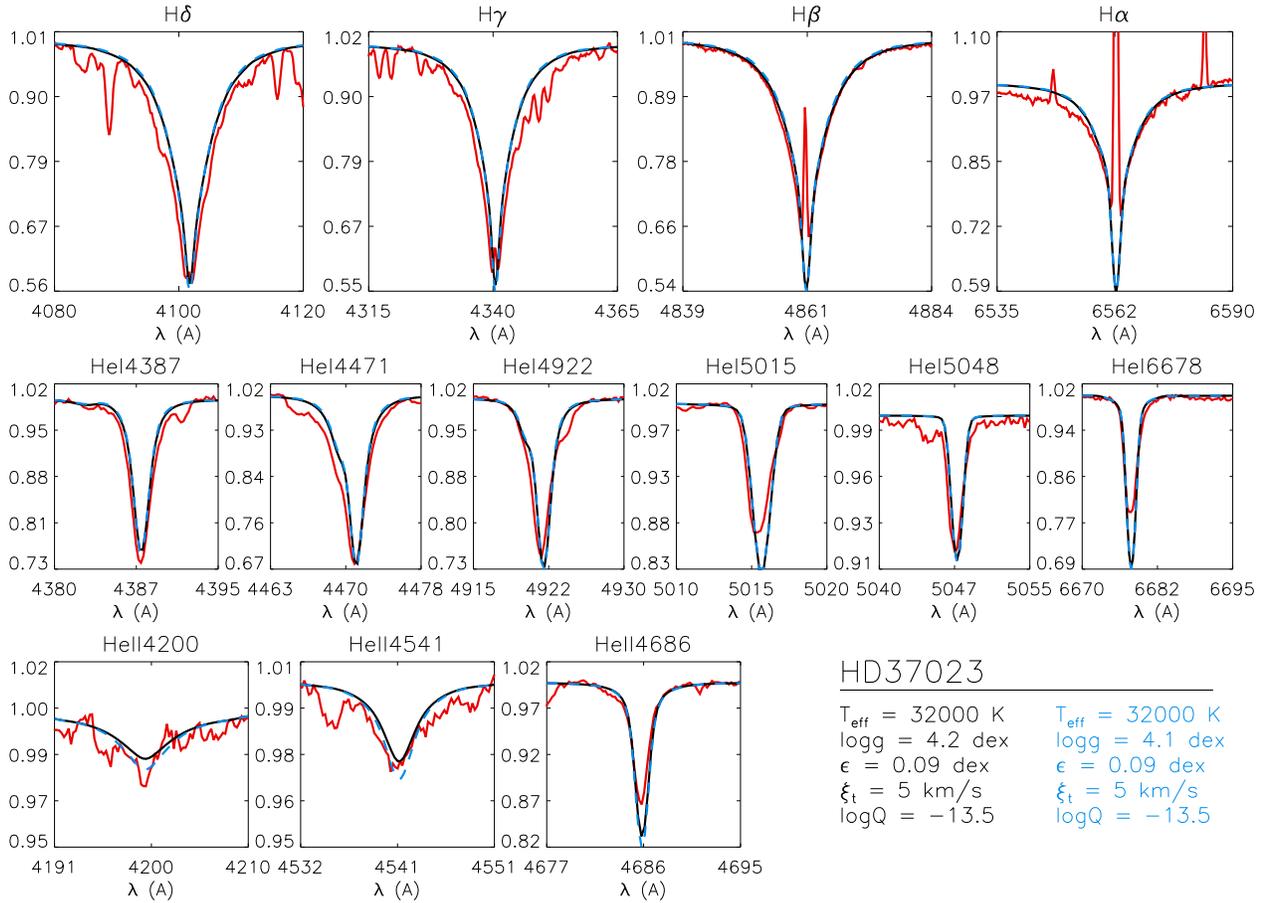}
\caption{As Fig. \ref{HD37020_fit} for HD\,37023 ($\theta^1$ Ori D, B0.5V). 
         A variation of 0.1 dex in \grav\ has been considered in this case.}
		 \label{HD37023_fit}
\end{figure*}

\begin{figure*}[ht!]
\centering
\includegraphics[width=12.cm,angle=90]{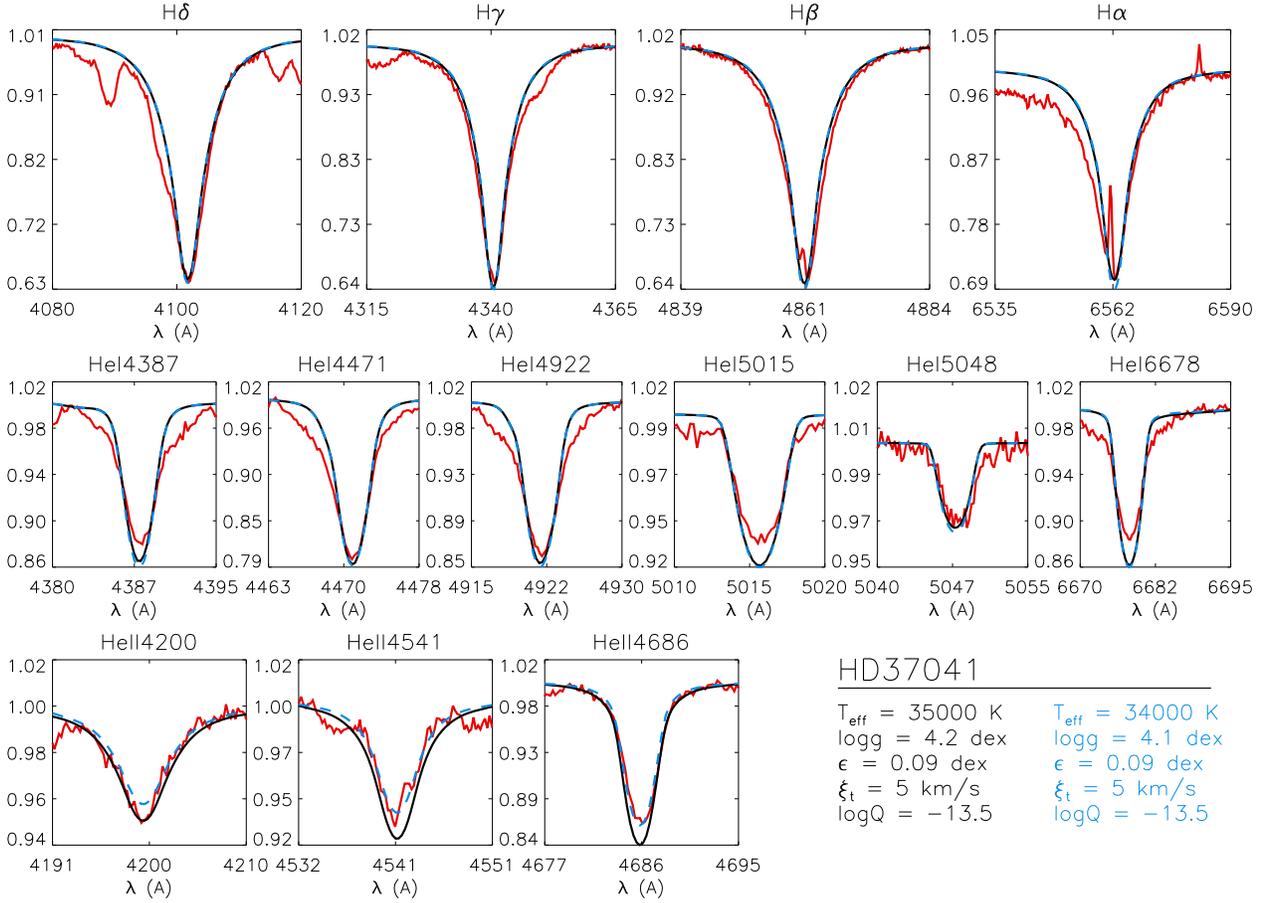}
\caption{As Fig. \ref{HD37020_fit} for HD\,37041 ($\theta^2$ Ori A, O9V). 
         A variation of 1000 K in \Teff\ along with 0.1 dex in \grav\ has 
		 been considered in this case.
		 Note that the broad wings of the \ion{He}{i} cannot be fitted
		 correctly (see text)}
		 \label{HD37041_fit}
\end{figure*}

\begin{figure*}[ht!]
\centering
\includegraphics[width=12.cm,angle=90]{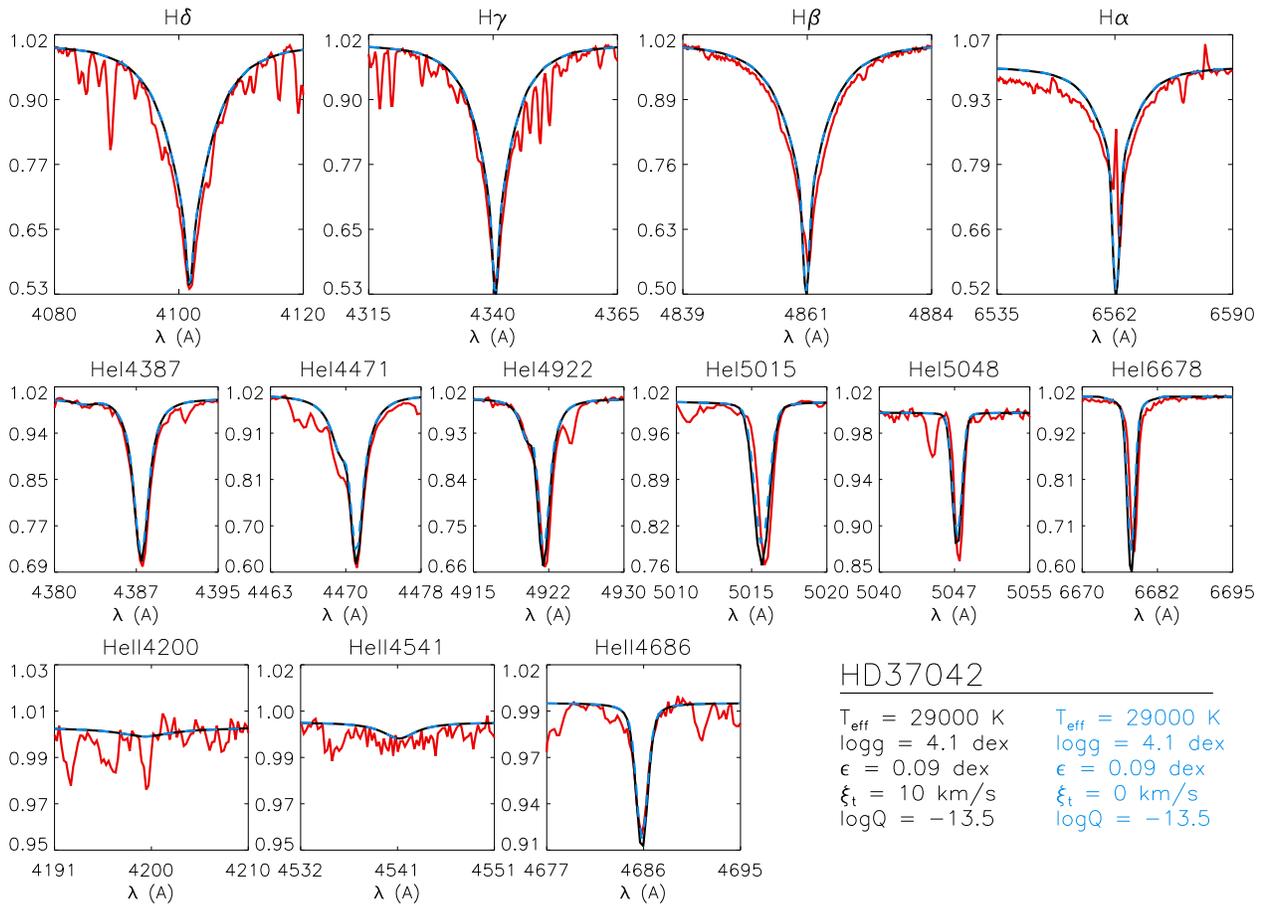}
\caption{As Fig. \ref{HD37020_fit} for HD\,37042 ($\theta^2$ Ori B, B0.5V). 
         For this star, the \ion{He}{i} lines fit
		 better if a microturbulence of 10 \kms\ is considered}\label{HD37042_fit}
\end{figure*}

\begin{figure*}[ht!]
\centering
\includegraphics[width=12.cm,angle=90]{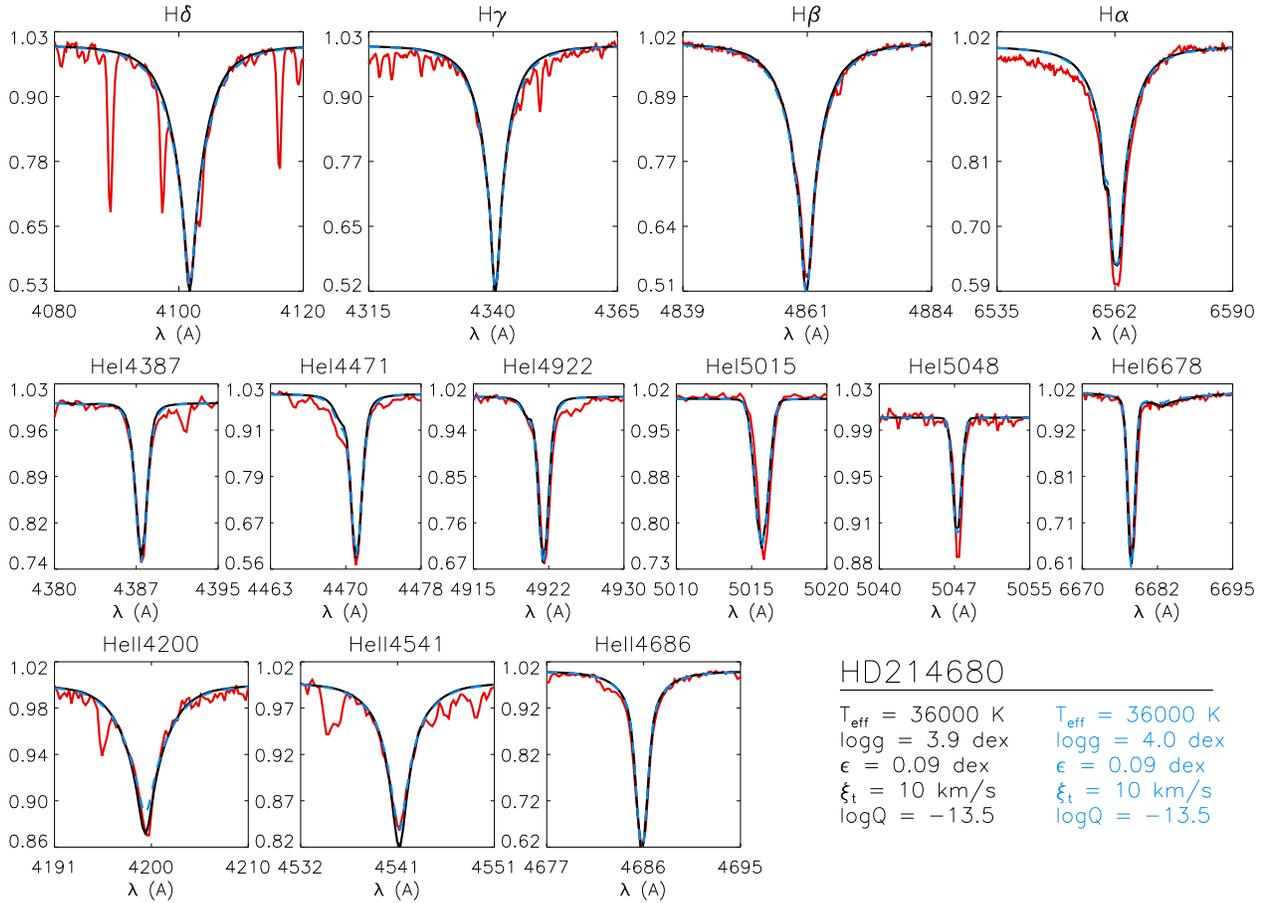}
\caption{As Fig. \ref{HD37020_fit} for HD\,214680 (10 Lac, O9V). 
         A variation of 0.1 dex in \grav\ has been considered 
		 in this case.}\label{10Lac_fit}
\end{figure*}

\begin{figure*}[ht!]
\centering
\includegraphics[width=12.cm,angle=90]{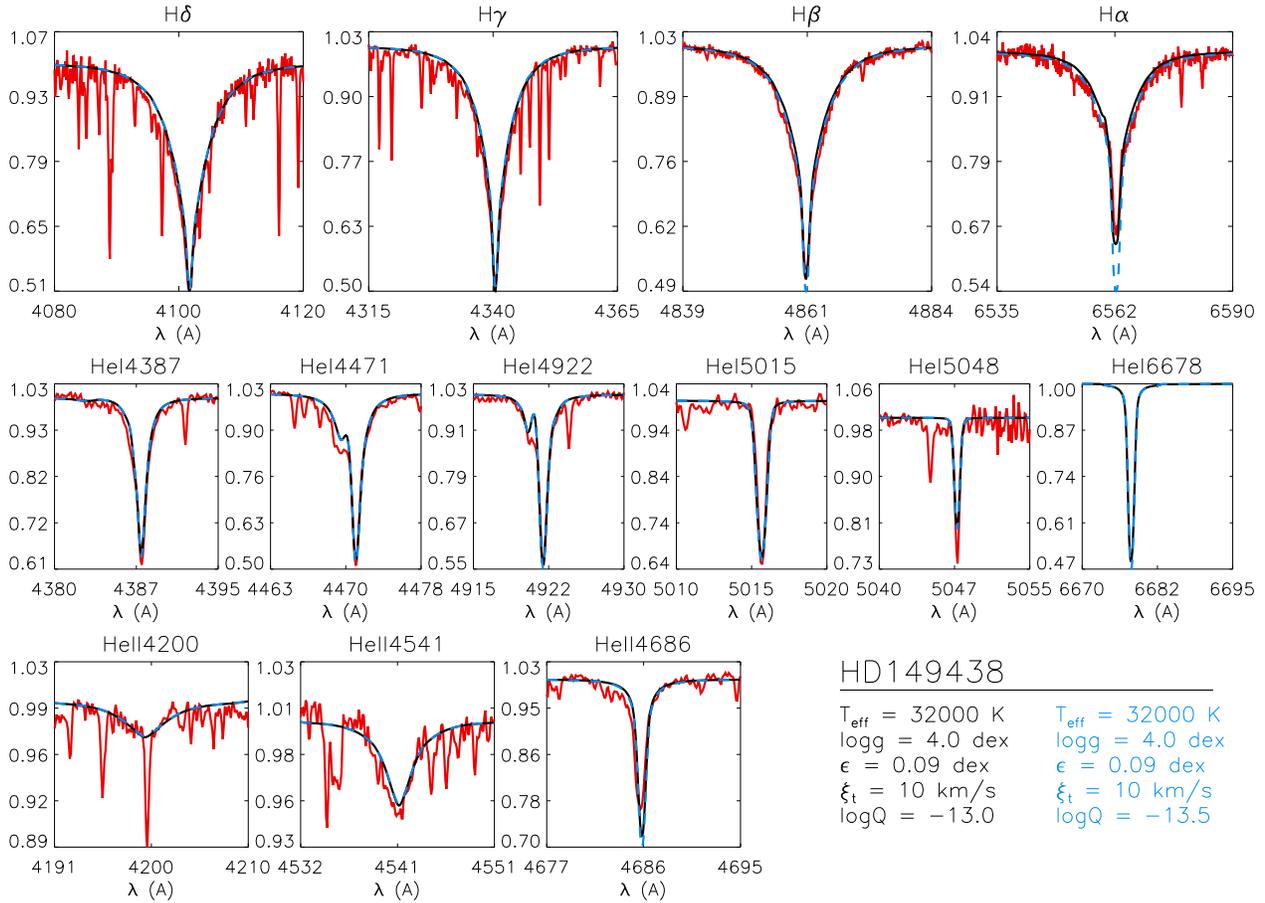}
\caption{As Fig. \ref{HD37020_fit} for HD\,149438 ($\tau$ Sco, B0.2V). 
         A variation of 0.5 dex in log\,$Q$ has been considered in 
		 this case. The H$_{\alpha}$ line is not contaminated by
		 nebular emission for this star and a more accurate value
		 of log\,$Q$ can be determined. The \ion{He}{i}\,$\lambda$ 6678
		 is out of the observed range in the {\sc caspec} spectrum
		 of $\tau$ Sco.}\label{tauSco_fit}
\end{figure*}

\begin{figure*}[ht!]
\centering
\includegraphics[width=12.cm,angle=90]{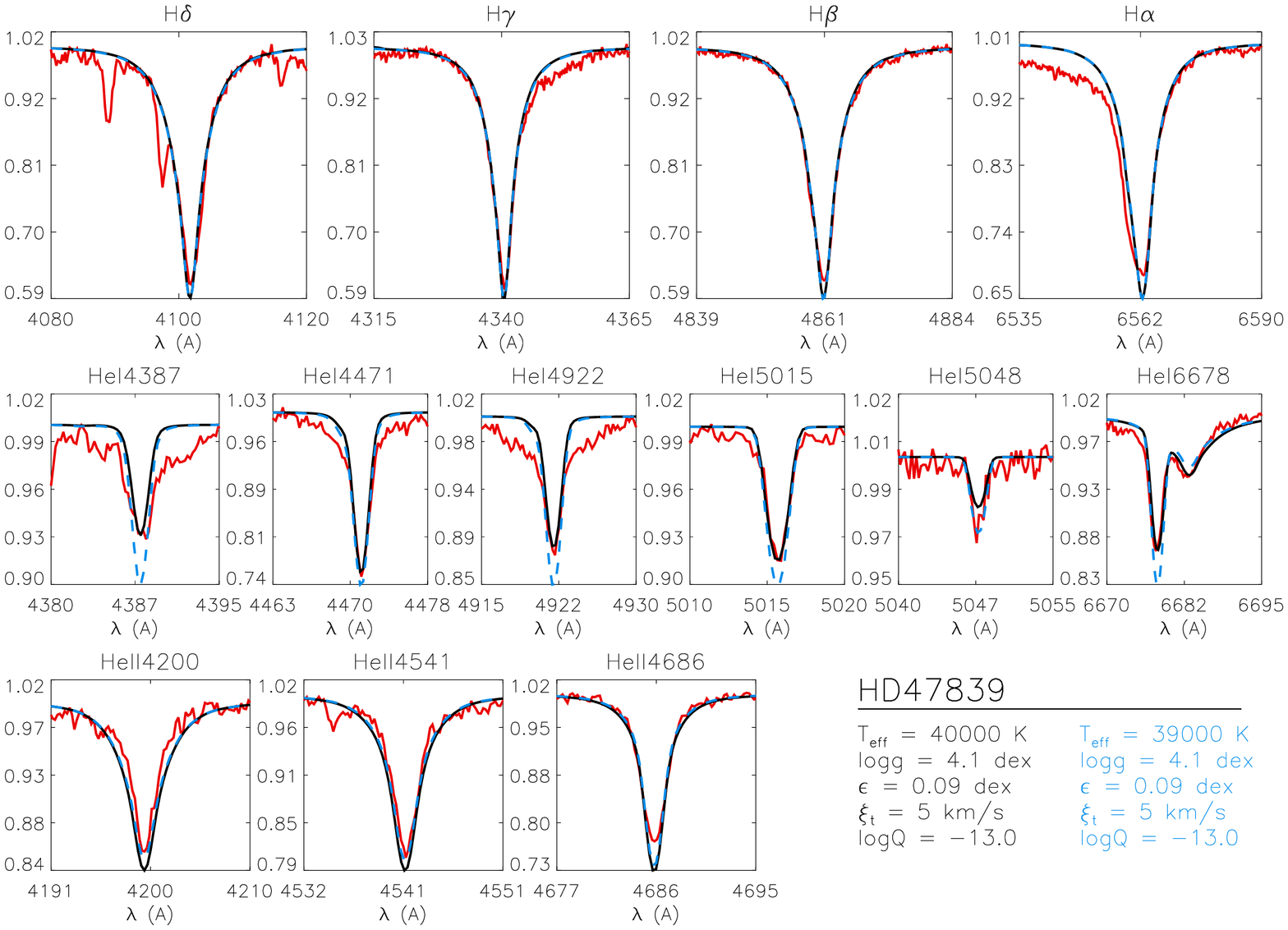}
\caption{As Fig. \ref{HD37020_fit} for HD\,47839 (15 Mon, O7V). 
         A variation of 1000 K in \Teff\ has been considered 
		 in this case.}\label{15Mon_fit}
\end{figure*}

\begin{table*}[!ht]
\begin{center}
\scriptsize{
\begin{tabular}{c c c c c c c c}    
\noalign{\smallskip}
\hline
\noalign{\smallskip}
HD\, & Name & \Teff\ & \grav\ & $R$ ($R_{\odot}$) & \mass 
(\mass$_{\odot}$) & log(\lum/\lum$_{\odot}$) & log\,$Q$\\
 & & $\pm$\,1000 K & $\pm$\,0.1 dex & & & & \\ 
\noalign{\smallskip}
\hline
\noalign{\smallskip}
HD\,37020  &  $\theta^1$ Ori A & 30000 & 4.0 &
6.3 $\pm$ 0.9 & 14 $\pm$ 5 & 4.45 $\pm$ 0.13 & $<$\,-13.5 \\
HD\,37022  & $\theta^1$ Ori C & 39000 & 4.1 & 
9.9 $\pm$ 1.4 & 45 $\pm$ 16 & 5.31 $\pm$ 0.13 & --- \\ 
HD\,37023  & $\theta^1$ Ori D & 32000 & 4.2 &
5.6 $\pm$ 0.8  & 18 $\pm$ 6 & 4.47 $\pm$ 0.13 & $<$\,-13.5 \\ 
HD\,37041  & $\theta^2$ Ori A & 35000 & 4.1 &
8.2 $\pm$ 1.1  & 39 $\pm$ 14 & 4.93 $\pm$ 0.13 & $<$\,-13.5 \\ 
HD\,37042  & $\theta^2$ Ori B & 29000 & 4.1 & 
4.5 $\pm$ 0.6  & 9 $\pm$ 3 & 4.11 $\pm$ 0.13 & $<$\,-13.5 \\ 
\noalign{\smallskip}
HD\,214680 & 10 Lac & 36000 & 3.9 & 
8.3 $\pm$ 1.1  & 20 $\pm$ 7 & 5.02 $\pm$ 0.13 & $<$\,-13.5 \\ 
HD\,47839 & 15 Mon & 40000 & 4.1 &
9.3 $\pm$ 1.3 & 40 $\pm$ 14 & 5.30 $\pm$ 0.13 & -13.0 \\  
HD\,149438 & $\tau$ Sco & 32000 & 4.0 &  
5.6 $\pm$ 0.8  & 11 $\pm$ 4 & 4.47 $\pm$ 0.13 & -13.0 \\  
\noalign{\smallskip}
\hline
\\
\end{tabular}
}
\rm
\caption{\footnotesize Stellar parameters derived from \fastwind\ 
analysis. Only an upper limit for log\,{\em Q} can be derived for 
these stars. The microturbulences considered for the HHe analysis in 
each star are shown in the corresponding fitting plots. A normal 
value for the He abundance has been considered for all the stars 
($\epsilon$\,=\,0.09).}\label{t3}
\normalsize
\end{center}
\end{table*}
The analyses have been performed using the latest version of \fastwind 
(an acronym for {\sc f}ast {\sc a}nalysis of {\sc st}ellar atmospheres with 
{\sc wind}s), a code which was originally described by Santolaya-Rey et 
al. (\cite{San97}). See Puls et al. (\cite{Pul05}) for the newest
description of the code along with a discussion of comparisons 
with previous models and other spherical mass-losing codes. The 
latest version uses a more complete {\em line-blanketing} and a 
temperature correction method based 
on the energy balance of electrons. The technique used for the 
derivation of the stellar parameters is already standard and has 
been described elsewhere (Herrero et al. \cite{Her02}; Repolust et 
al. \cite{Rep04}). We only give here the main points.
The analyses are based on visual fitting of hydrogen Balmer lines 
and \ion{He}{i} and \ion{He}{ii} lines. Through the \ion{He}{i}/\ion{He}{ii} 
ionization equilibrium, the effective temperature can be estimated; 
the wings of the Balmer lines are useful for the determination of 
the gravity and can give us some information about the stellar 
wind.\\
The code also needs other parameters, such as microturbulence, 
He abundance and wind properties (mass loss, terminal velocity 
and $\beta$ parameter). Actually wind properties are related 
through the $Q$ parameter ($Q$\,=\,$\dot{M}$\,/\,(\vinf\radius)$^{1.5}$).\\
\newline
Once the observed lines are fitted with the modeled ones, effective temperature, 
gravity, He abundance, microturbulence and log\,$Q$\ are determined. The
low density in the winds of the studied objects makes the spectrum insensitive
to changes in $Q$, so that we can only determine upper values in most cases.
Microturbulence has no effect on the H/He spectrum of early type stars of large
gravity, as it has been shown by Villamariz \& Herrero (\cite{Vil02}). Therefore
only effective temperatures, gravities and He abundances are determined for this
step of the analysis. Of course, microturbulence is important for the derivation
of metallic abundances and will be determined in the corresponding section.
The code also provides the emergent flux distribution and then mass, 
radius and luminosity can be calculated if $M_{\rm v}$ is known (see Herrero 
et al. \cite{Her92}).\\
\newline
Errors in \Teff\ and \grav\ can be established generating a microgrid 
around central values; visual comparisons between modelled lines and 
observations allow us to determine the range of possible values for 
these parameters. For further comments on the effects of varying 
the various physical parameters used in the analyses, their mutual 
interplay and their error bars see viz. Repolust et al. (\cite{Rep04}),
Herrero et al. (\cite{Her02}), Villamariz \& Herrero (\cite{Vil00}), and
Villamariz et al. (\cite{Vil02}).
Errors in \radius, \mass\ and $L$ are calculated considering the
propagations of the uncertainties in \Teff, \grav\ and $M_{\rm v}$.\\
\newline
The fits of the synthetic \fastwind \ion{H}{} and \ion{He}{i-ii} profiles 
to the observed ones are shown in Figures \ref{HD37020_fit} to
\ref{15Mon_fit}. The derived parameters for our sample of stars 
are shown in Table \ref{t3}, corresponding to the best fits.\\
Some comments on 
the individual analyses and the comparison between spectroscopic and 
evolutionary results are given in Sects. \ref{individual} and \ref{discuss}.  
\subsection{Comments on the Trapezium cluster stars}\label{individual}
\begin{description}
\item{\em HD\,37020:}
This is an eclipsing binary discovered by Lohsen (\cite{Loh75}). The period 
is $P$\,=\,65.43233 d, the magnitude range $m_{\rm v}$\,=\,6.7\,-\,7.6 and the
eclipse lasts $\sim$\,20 h. The primary is a B0.5V star which has many 
isolated metal lines. We have not found lines of the secondary
star in our spectrum. This fact, together with the 
characteristics of the binary system described before, suggest that the 
secondary is a cooler, smaller and less luminous star, and that the change
in the visual magnitude is due to the eclipse of the primary by its companion.
Therefore the optical spectrum is mainly dominated by the B0.5V star.\\
\newline
The Fourier method has been applied to some \ion{O}{ii}, \ion{Si}{iii-iv}, 
\ion{N}{ii} and \ion{He}{i} lines, deriving a \vsini\,=\,55\,$\pm$\,0.6 
\kms. Metal and \ion{He}{i} lines are in agreement. The \vsini\ derived 
through the linewidth measurement method is consistent with this value (see 
Table \ref{t2}).\\
\newline
Figure \ref{HD37020_fit} shows the good fit of the \fastwind profiles to 
the observed spectra for the parameters given in Table \ref{t3} (except 
for the forbidden component of \ion{He}{i}\,$\lambda$4471 which is not well
reproduced throughout our analyses. Note also that the apparent bad fit of 
\ion{He}{ii}\,$\lambda$4200 is due to the blend with the \ion{N}{iii} line
at the same wavelength). 
We are very close to the applicability limit of the \ion{He}{i-ii} ionization 
equilibrium for deriving the \Teff\ as \ion{He}{i}\,$\lambda$4200 and
\ion{He}{i}\,$\lambda$4541 lines are very faint; however these lines are still 
sufficiently sensitive to changes in \Teff\ and \grav\ for deriving the 
stellar parameters accurately. 
The stellar parameters obtained here are in very good agreement with those 
obtained by Cunha \& Lambert (\cite{Cun92}) using the Str\"omgren index 
$c_0$ and the wings of H$_{\gamma}$ with used Kurucz's 
(\cite{Kur79}) LTE model atmospheres (this is not the case for the other 
two stars in common with these authors). A comparison of values is given 
in Table \ref{Cunhatable}.\\
\item{\em HD\,37022:} This is the main ionising source of the Orion nebula. 
See section \ref{HD37022} for a detailed study of this star.\\
\begin{table}[!ht]
\begin{center}
\scriptsize{
\begin{tabular}{c c c}    
\noalign{\smallskip}
\hline
\noalign{\smallskip}
 Star & \Teff\ (K) & \grav\ (dex) \\
\noalign{\smallskip}
\hline
\noalign{\smallskip}
HD\,37020 & 29970 / 30000 & 3.92 / 4.0 \\
HD\,37023 & 32600 / 32000 & 4.70 / 4.2 \\
HD\,37042 & 31600 / 29000 & 4.70 / 4.2 \\
\noalign{\smallskip} 
\hline
\\
\end{tabular}
}
\normalsize
\rm
\caption{\footnotesize Comparison of stellar parameters for 
          HD\,37020, HD\,37023, HD\,37042. First values refer 
		  to Cunha \& Lambert (\cite{Cun92}) determinations,
		  second values to this work. We see that there is 
		  excellent agreement for HD\,37020, but poor agreement
		  (specially for \grav) for the other two stars.
\label{Cunhatable}
}
\end{center}
\end{table}
%
\item{\em HD\,37023:} This is the only star in the four Trapezium cluster
stars ($\theta^1$ Ori) without a binary companion (Preibisch et al. 
\cite{Pre99}). Robberto et al. (\cite{Rob04}) find indications that this 
star is surrounded by a photoevaporated circumstellar disk.\\
The Fourier method gives a \vsini\,=\,49.0\,$\pm$\,0.9 \kms\ for this star. 
Again, there is agreement with the linewidth measurement method.\\
\newline
Figure \ref{HD37023_fit} shows the fitting of the HHe lines. Observed 
\ion{He}{i} lines are slightly broader than the theoretical ones. 
Table \ref{Cunhatable} shows the comparison between the stellar
parameters we have derived and those by Cunha \& Lambert (\cite{Cun92}).
In this case the agreement is not so good as for HD\,37020, although
the \Teff\ are compatible, the \grav\ they derived is very high.\\
\item{\em HD\,37041:} This is a single-lined spectroscopic binary
(see Howarth et al. \cite{How97}). These authors find a single peak in
the cross correlation function of the IUE spectrum, indicating that
the spectrum of the primary is not seriously contaminated.\\
\newline
Comparing the spectrum of this O9V star with that
of the standard star 10 Lac (also classified as O9V) we have found that 
there are no unblended metal lines due to its high rotational velocity
(except \ion{Si}{iv}\,$\lambda$4089, but it is in the blue wing of
H$_{\delta}$). A good \vsini\ determination has been possible using the 
Fourier method with the \ion{He}{i} lines. A \vsini\,=\,131\,$\pm$\,4 
\kms\ has been derived. We have used the \ion{Si}{iv}
line for checking the reliability of this value; a \vsini\,=\,136\,$\pm$\,5 
\kms\ is obtained. \ion{He}{ii} does not give good results.
A \vsini\,=\,140\,$\pm$\,11 \kms\ is derived using the \fwhm method 
for the \ion{He}{i}\,$\lambda$5015 line.\\
\newline
Figure \ref{HD37041_fit} shows the fitting of the synthetic profiles with
the observed ones. A \vsini\,=\,131 \kms\ has been considered for the 
\ion{H-He}{} analysis. The wings of the \ion{He}{i} lines cannot be 
well fitted; this might be explained by the presence of a companion. \\
\item{\em HD\,37042:} We have found good agreement between metal and 
\ion{He}{i} lines when using the Fourier method. The \vsini\ derived is 
34.0\,$\pm$\,0.5 \kms. The value we obtain with the \fwhm method is 
36\,$\pm$\,3 \kms, in agreement with the former one.\\
\newline A very good fit of the observed and synthetic \fastwind
profiles is obtained (see Figure \ref{HD37042_fit}). Again the 
forbidden component of \ion{He}{i}\,$\lambda$4471 is too weak. For
this star, the \ion{He}{i} lines fit better if a microturbulence
of 10 \kms\ is considered.\\
\newline
This is the third star in common with Cunha \& Lambert (see Table 
\ref{Cunhatable}); we also find for this star (as for HD\,37023) that 
the stellar parameters derived by these authors are very different 
from ours (they obtain a very high \grav\ and a \Teff\,$\sim$\,3000 
K higher).
\end{description}
\subsection{Comments on the reference stars}\label{individual2}
\begin{description}
\item{\em HD\,214680:} This star has been previously considered as
standard for stellar parameter determination through the \ion{H-He}{} 
analysis (see Herrero et al. \cite{Her02}
and references therein). We have reanalysed here the spectrum of the
star, obtained from a new observing run covering a larger spectral 
range (the \ion{He}{i}\,$\lambda\lambda$5015, 5048, 6678 and 
\ion{He}{ii}\,$\lambda$6683 lines can be used for the new analysis).
We are using a new \fastwind version (Puls et al. \cite{Pul05}), 
slightly different from that used by Herrero et al. The new 
analysis will be useful for checking the consistency of the latest
version of the code in this stellar parameters regime.\\
\newline
The projected rotational velocity of this star has been determined 
accurately by means of the \fwhm method (\vsini\,=\,37\,$\pm$\,4 \kms). 
The Fourier method applied to the \newton+\ids\ spectrum gives a 
\vsini\,=\,30\,$\pm$\,0.8 \kms. This larger difference could be due 
to the fact that the \vsini\ is close to 
the computational Fourier transform limit ($\sim$20 \kms\ for this 
spectrum), or because the microturbulence is affecting the determination 
of the rotational first zero in the Fourier transform (Gray \cite{Gra73}).\\
\newline
The whole set of HHe lines are perfectly fitted with the \fastwind
synthetic profiles (Figure \ref{10Lac_fit}). The parameters derived by 
Herrero et al. (\cite{Her02}) are \Teff\,=\,35500 K, \grav\,=\, 3.95 and 
$\epsilon$ = 0.09. Our results are in agreement with those values.\\
\item{\em HD\,149438:} This star was selected for comparison in the
stellar oxygen abundance analysis of the B0.5V Trapezium cluster 
stars (see section \ref{abundance}). It has been studied elsewhere 
(Martin \cite{Mar04}; Przybilla \& Butler \cite{Prz04}; Kilian 
\cite{Kil94}; Sch\"onberner et al. \cite{Sch88}; Becker \& Butler 
\cite{Bec88}; 
Peters \& Polidan \cite{Pet85}; Kane et al. \cite{Kan80}). Comparing 
isolated oxygen and silicon lines to a set of rotationally broadened 
profiles, Kilian et al. (\cite{Kil91}) obtained a \vsini\ = 19 \kms\ 
for this star. The Fourier method applied to the {\sc caspec} spectrum only 
allow us to say that the \vsini\ is lower than 13 \kms; this is because
in this case the effect of the microturbulence over the broadening of the 
metal lines can be comparable with that produced by the rotation, so the
first zero could be associated with the microturbulence instead of the
\vsini\ (Gray \cite{Gra73}).\\
\newline
A very good fit with the synthetic \fastwind profiles is obtained for this
star (see Figure \ref{tauSco_fit}). In this case the problem with the 
forbidden component of the \ion{He}{i}\,$\lambda$4471 can be clearly seen.
Table \ref{t_comp_tau} (Sect. \ref{tausco}) summarized the stellar 
parameters obtained in this and previous work.\\
\item{\em HD\,47839:} This star was selected for comparison with the
main ionizing source in Orion ($\theta^1$ Ori C). Some comments on
the analysis of this star and the comparison with $\theta^1$ Ori C
are presented in section \ref{HD37022}. It was classified as
O7V(f) by Walborn (\cite{Wal72}). Gies (\cite{Gie93}) pointed for the 
first time that this star is a spectroscopic binary with a period of 
25 years. He estimated a mass for the primary of \mass\,=\,34 
\mass$_\odot$, and suggested that the secondary has a spectral type 
O9.5Vn (with very broad lines).\\
We have used the spectrum of HD\,214680 (O9V) convolved with a high 
\vsini\ ($\sim$\,350 \kms), for recognizing lines in the spectrum of 
HD\,47839 not contaminated by the secondary star; three metal lines 
were found. Using these lines (\ion{Si}{iv}\,$\lambda\lambda$\,4212, 
4654 and \ion{N}{iii}\,$\lambda$ 4379) a \vsini\,=\,67\,$\pm$\,4 \kms\ 
has been determined. A similar value has been derived applying the 
\fwhm method to the same lines (66\,$\pm$\,6 \kms).\\
\newline
The fitting of the H and He lines for the stellar parameters
show how the \ion{He}{i} are contaminated by the secondary star lines. The 
spectroscopic derived mass (36 $\pm$ 9 \mass$_{\sun}$) is in very good 
agreement with the dynamical mass derived by Gies (\cite{Gie93}). Herrero 
et al. (\cite{Her92}) derived a \Teff\,=\,39500 K, \grav\,=\,3.70 and 
$\epsilon$\,=\,0.07 for this star. Although we would expect to obtain a lower 
\Teff\ due to the inclusion of {\em line-blanketing} effects, the value we 
have obtained is slightly higher because there is also a large difference between 
the \grav\ we derived (4.0) and the one by Herrero et al. (3.7). There is 
also another difference; we do not need a lower \ion{He}{} abundance for 
fitting the \ion{He}{} lines. This could be due to a binarity effect; when 
a composite spectrum is considered in a binary system, the lines can appear 
diluted or magnified due to the combination of the fluxes of the primary 
and the companion. If the system is out of eclipse the total flux will be 
higher than when the primary is observed isolated, so when the spectrum is 
normalized all the lines will appear diluted (and then a lower \ion{He}{} 
abundance is needed to fit the \ion{He}{} lines and a lower gravity is derived).
\end{description}
\subsection{Results of the stellar parameters study}\label{discuss}
\begin{figure}[ht!]
\centering
\includegraphics[width=6.6cm,angle=90]{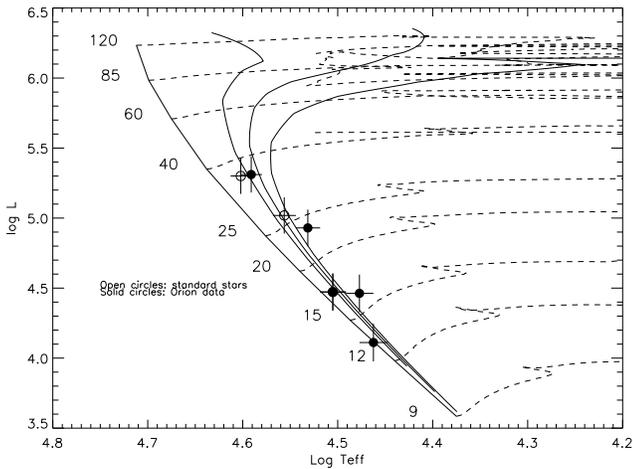}
\caption{HR diagram with the studied stars. Evolutionary tracks from
Meynet \& Maeder (\cite{Mey03}). Isocrones from Schaller et al.
(\cite{Sch92}), corresponding to 2, 2.5 and 3 Myr}\label{HR}
\end{figure}

From the optical spectra of the Orion stars only upper limits for the 
$Q$ parameter can be achieved. These estimations are based on
the effect of the wind on the \ion{He}{ii}\,$\lambda$4686 and 
H$_{\rm \alpha}$ lines (the later one is contaminated by the nebular 
contribution). Some tests have shown that the other \ion{H}{} and
\ion{He}{} lines are not affected by the uncertainties in log\,$Q$
for the range of values considered, so the derived parameters
will not be affected.\\
\newline
Masses, radii and luminosities have been derived for all the studied
targets (they are indicated in Table \ref{t3} together with their 
uncertainties). The main source of uncertainty for these parameters
is that associated with the absolute magnitude (except for very large
uncertainties in \grav). An error in $M_{\rm v}\sim$\,0.3 propagates 
to the mass, radius and logarithmic luminosity, giving uncertainties
of $\sim$\,37\,\%, 13\,\% and 3\,\% respectively.\\
\newline
The stars have been plotted on the HR diagram in Figure \ref{HR}. 
Evolutionary tracks from Meynet \& Maeder (\cite{Mey03}),
corresponding to initial masses ranging from 9 to 120 \mass$_{\odot}$
and initial rotational velocities of 0 \kms\ are also plotted. 
All stars are found in the Main Sequence close to
the ZAMS, as it is expected because of their youth. Nevertheless, we can
see the separation from the ZAMS increasing with luminosity, as pointed
out by Herrero et al. (\cite{Her04}). The loci of the Orion stars is consistent with an isochrone at about 
2.5$\pm$0.5 Myr, derived from the tracks with zero initial rotational 
velocity, which is slightly older than the upper limit given by 
Palla \& Stahler  (\cite{Pal99}, 2 Myr) and somewhat larger than 
other Trapezium age determinations found in the literature (e.g. 
Hillenbrand, \cite{Hile97}, $\leq$\,1 Myr). However, it has to be considered 
that, at large initial rotational velocities, the {\sc zams} is 
slightly shifted to the right and modifies the derived ages. Hence, 
until the role of the initial rotational velocities is properly 
understood (for example, the distribution of initial rotational 
velocities in clusters), the use of isochones for massive stars in 
very young clusters should be regarded with special caution.\\
\newline
A good agreement between gravities derived from the evolutionary 
tracks and those obtained from quantitative analysis of the spectra 
is found (see Table \ref{t_masas}). There is a trend for the most massive stars to have larger 
spectroscopic than evolutionary masses, but the number of objects is 
too small to draw any general conclusion.\\
\begin{table}[!h]
\begin{center}
\scriptsize{
\begin{tabular}{c c c c c}    
\noalign{\smallskip}
\hline
\noalign{\smallskip}
 Star & $M_{\rm spec} $ ($M_{\sun}$) &  $M_{\rm evol} $ ($M_{\sun}$) &
 log\,$g\,_{\rm spec}$ & log\,$g\,_{\rm evol}$\\
\noalign{\smallskip}
\hline
\noalign{\smallskip}
 $\theta^1$ Ori A   & 14 &  15  & 4.0 &  4.03 \\
 $\theta^1$ Ori C   & 45 &  33  & 4.1 &  3.98 \\
 $\theta^1$ Ori D   & 18  & 16  & 4.2 &  4.16 \\
 $\theta^2$ Ori A   & 39  & 22  & 4.1 &  3.97 \\
 $\theta^2$ Ori B   & 9  & 13  & 4.1 &  4.26 \\
 10\,Lac                & 20 &  25 &  3.9 &  4.02 \\
 15\,Mon               & 40 & 34  & 4.1  &  4.05\\
 $\tau$\, Sco        & 11 &  16 &  4.0  &  4.16 \\
\noalign{\smallskip} 
\hline
\\
\end{tabular}
}
\normalsize
\rm
\caption{\footnotesize Comparison of masses and gravities derived from the
evolutionary tracks and from the quantitative analysis of the
spectra. The quoted log\,$g_{\rm evol}$ values are given corrected to two decimal places
to be consistent with the corresponding evolutionary masses. Note however that these are not an indication
of the precision of these values, which we consider to be 0.1 dex.
\label{t_masas}
}
\end{center}
\end{table}
%
%
\section{Modeling the main ionizing star of the Orion nebula}\label{HD37022}
%
HD\,37022 ($\theta^1$ Ori C, O7V) is the main ionizing source of the Orion
nebula. We want to derive its stellar parameters as a first step to 
determine the effect of its ionizing flux on the photoionization of the
surrounding nebula in a consistent way. Once these parameters are known,
the spectral energy distribution can be modeled by using model atmosphere
codes. In this way one of the inputs used in the photoionization codes will
be fixed consistently.
\subsection{A historical review}
This star is known to have variable spectral features varying 
in phase or antiphase with a period of 15.422 $\pm$ 0.002 d (Stahl et al. 
\cite{Sta93}; Walborn \& Nichols \cite{Wal94}; Stahl et al. \cite{Sta96}). 
These variable features were discovered after Conti (\cite{Con72}) showed 
for the first time that $\theta^1$ Ori C has a variable inverted P-Cygni 
profile in the \ion{He}{ii} 4686 line. Among 
them are H$_{\alpha}$ emission, variability in the equivalent 
width of some atmospheric and wind lines, and X-ray emission (Caillault 
et al. \cite{Cai94}; Gagn\'e et al. \cite{Gag97}).
Different explanations for this variability were postulated.
The possibility of $\theta^1$ Ori C being a binary and this binarity 
explaining the spectral variability has been dismissed (Stahl et al. 
\cite{Sta96}). The variability has been associated with the rotation 
of the star.
Stahl et al. (\cite{Sta96}) proposed the presence of a dipolar magnetic 
field in $\theta^1$ Ori C, with the magnetic pole inclined 45$^{\circ}$ 
from the rotation axis (which is inclined 45$^{\circ}$ from the line of 
sight). The geometry of this system would imply alignment between 
magnetic pole and the line of sight at phase 0.5 and 
they would be perpendicular at phase 0.0 (when maximum H$_{\alpha}$ 
emission is found). Babel \& Montmerle (\cite{Bab97}) proposed the 
magnetically confined wind-shock model (MCWS) for explaining the variability 
in the star. According to this model, the radiatively line driven wind is 
confined by a dipolar magnetic field towards the magnetic equator of the 
system, generating a cold, dense disk due to the collision of material 
coming from both hemispheres, and a hot post-shock region.\\
\newline
The wind characteristics of $\theta^1$ Ori C were determined by Howard \& Prinja 
(\cite{How89}) and Stahl et al. (\cite{Sta96}) through UV spectrum studies. 
The former derived a mass loss rate of 4\,x\,10$^{-7}$ \mass$_{\sun}\,$yr$^{-1}$, 
the latter through the absorption in \ion{C}{iv} lines, determined a terminal 
velocity somewhat greater than 2500 \kms. It is common that O7V stars have 
stellar winds; what is not so common is the detection of the presence of magnetic 
fields in O stars. Donati et al. (\cite{Don02}) succeeded in the detection 
of Zeeman features in the spectrum of $\theta^1$ Ori C through 
spectropolarimetric observations with the Anglo-Australian Telescope. They 
detected variability in 
the Stokes V profiles of some photospheric metal lines. This variability 
has a coherent modulation with the period derived from other variable 
features. However, the geometry derived was in contradiction with that from 
Stahl et al. (\cite{Sta96}), with the magnetic pole aligned with the line 
of sight at phase 0 (they found a maximum in the longitudinal component of the 
magnetic field at this phase).
\subsection{Preliminary study of the spectrum}\label{variable}
\begin{figure*}[ht!]
\centering
\includegraphics[width=6.5cm,angle=90]{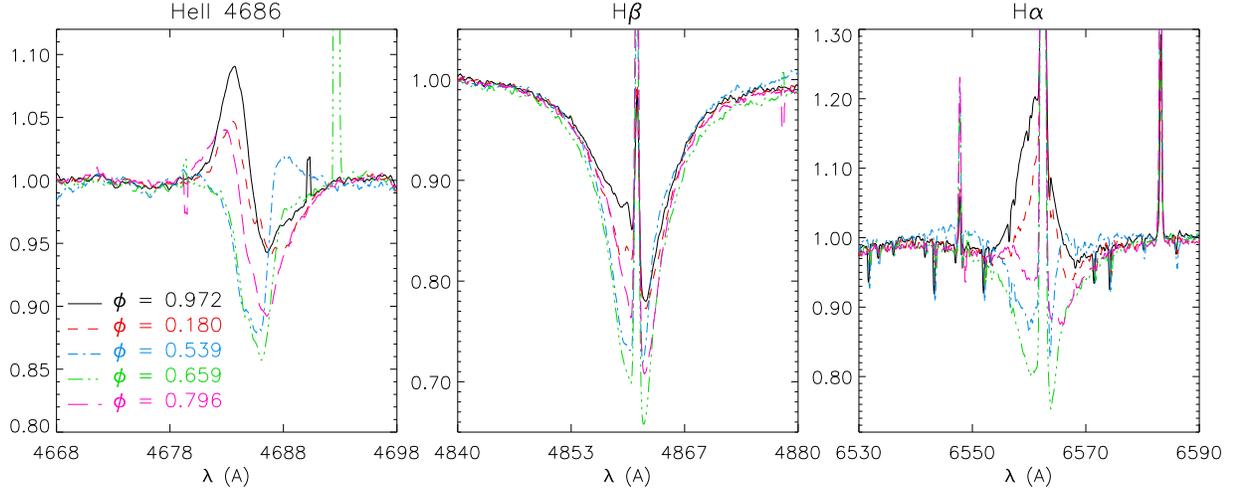}
\caption{The most representative phases have been selected for showing 
         the variability of the \ion{He}{ii}\,$\lambda$4686, 
		 H$_{\rm \beta}$ and H$_{\rm \alpha}$ lines in the 
		 $\theta^1$ Ori C spectrum. The variable feature associated
		 with the inverted P-Cygni profile discovered by Conti in the
		 \ion{He}{ii}\,$\lambda$4686 line is also present in the hydrogen 
		 Balmer lines. Another emission feature can be clearly seen in
		 the red wing of these lines at phase $\sim$\,0.5. The narrow
		 emission features in the Balmer lines are nebular lines.
 }\label{OriCvarlines}
\end{figure*}

\begin{figure}[t!]
\centering
\includegraphics[width=6.0cm,angle=90]{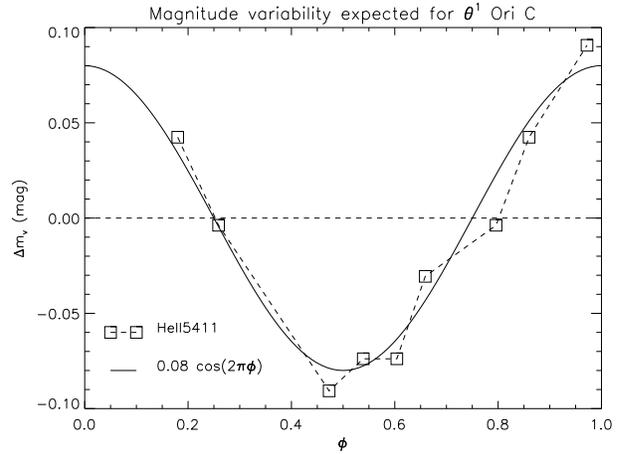}
\caption{Magnitude variability expected from the study of the 
         \ion{He}{ii}\,$\lambda$5411 line. The other \ion{He}{i-ii}
		 lines follow a similar behavior. The presence of a disk could 
		 be responsible of this variability (see text for details). 
		 The solid line corresponds to a sinusoidal curve with a 
		 maximum change in m$_{\rm v}$ of 0.16, presented for 
		 comparison.}
		 \label{magOriC}
\end{figure}

Having spectral variability, it is very important to understand the 
cause of this variability and to determine which lines are reliable 
for the stellar atmosphere modeling before comparing synthetic and 
observed \ion{H}{}\,-\,\ion{He}{} profiles. 
Preliminary work with the \newton+\ids\ spectrum showed that a better 
spectral resolution was needed to apply the Fourier method for obtaining 
the \vsini. This spectrum did not allow us to have neither a good enough
sampling ($\Delta\lambda$), nor to carry on a study of the variability, 
so we decided to use some \feros\ spectra with better quality and covering 
all variability phases (see Table \ref{tvariable}), that are available
in the {\sc eso} archive.\\ 
\newline
Through the study of the \feros\ spectra observed in the different phases
we have found all the variable spectral features described in Stahl et al. 
(\cite{Sta96}) and some new ones:
\begin{itemize}
\item {\bf \ion{He}{ii} 4686: } Broad emission appears in the blue wing 
of the line (the so-called inverted P-Cygni profile, with 
maximum at $\phi$\,$\sim$\,0). Broad emission is also present in the red 
wing (maximum at $\phi$\,$\sim$\,0.5, minimum at $\phi$\,$\sim$\,0.8).
\item {\bf Balmer lines: } These lines are affected by the same broad 
emission features than those in \ion{He}{ii}\,$\lambda$4686. The emission
features are stronger in H$_{\rm \alpha}$ and H$_{\rm \beta}$ than in the
other Balmer lines
\item{{\bf Metal and \ion{He}{i-ii} lines: } All the line strength varies 
in phase, being larger at $\phi$\,$\sim$\,0.}
\end{itemize}
\begin{table}[h]
\begin{center}
\scriptsize{
\begin{tabular}{c c c c}    
\noalign{\smallskip}
\hline
\noalign{\smallskip}
 Name & Date & $MJD$-2.400.000,5 & $\phi$  \\
\noalign{\smallskip}
\hline
\noalign{\smallskip}
 f07341+51 & 16/10/98 & 51102.31 & 0.180 \\
 f85221    & 26/07/99 & 51385.43 & 0.539 \\
 f03551+61 & 08/10/98 & 51094.28 & 0.659 \\
 f04711+21 & 10/10/98 & 51096.39 & 0.796 \\
 f15241    & 28/11/98 & 51145.37 & 0.972 \\
\noalign{\smallskip}
\hline
\\
\end{tabular}
}
\normalsize
\rm
\caption{\footnotesize \feros\  spectra used for the study of the spectral
         variability of $\theta^1$ Ori C. All spectra have been downloaded 
		 from the {\sc eso}-\feros\  database except 
		 f85221, 
		 kindly 
		 provided by O. Stahl. The different phases have been calculated 
		 from $MJD_0$ - 2.400.000,5 = 48832,5 (Stahl et al. \cite{Sta96}) and 
		 $P$\,=\,15.422 days}\label{tvariable}
\end{center}
\end{table}
This variability can be easily explained considering the model 
proposed by Stahl et al. (\cite{Sta96}) and developed by Babel 
$\&$ Montmerle (\cite{Bab97}). According to this model we would 
have an O7V star with a disc. The disc is produced by the confinement 
of the wind by a dipolar magnetic field through the magnetic equator. 
We will have a cool disc with material falling back to the stellar surface. 
If we consider that at phase 0 the cool disc is seen edge-on, the
blue shifted emission appearing in \ion{He}{ii}\,$\lambda$4686,
H$_\alpha$ and the other hydrogen Balmer lines can be explained as
stellar photons absorbed and reemited with a doppler shift corresponding 
to the velocity of the disc material falling to the surface of the 
star (in a process
similar to what occurs in a stellar wind but with blue-shifted emission
and red-shifted absorption). As density in the disc is very high
then a strong blue-shifted emission will appear. At phase 0.5, 
when the disc is seen face-on the blue-shifted emission disappears.
The emission appearing in the red wings of the former lines could be
explained as the effect of the scattering of stellar photons by the 
wind material confined by the magnetic field and accreting to the disc
(see figure \ref{OriCvarlines}).
The disc will also have continuum emission that will affect the 
total continuum flux received from the star. This effect will be 
maximum when the disc is face-on because the emitting 
region is larger at this phase. The variability observed in 
\ion{He}{i}, \ion{He}{ii} and metal lines (except for the emission
in \ion{He}{i}\,$\lambda$4686) is only a consequence 
of this effect. As the total continuum flux is varying with the 
phase, the normalized spectrum will be affected. All absorption lines 
will be artificially weaker when the disc continuum emission is at 
maximum. This variability in the lines can allow us to estimate 
the amount of continuum flux that comes from the disc, and then 
the visual magnitude variability. By assuming that at phase 0.0 the
lines are not affected by the continuum emission from the disc, we have
scaled the spectra at the other phases to fit in the former spectrum;
the scaling factor will be related to the ratio of visual fluxes (i.e.
the difference in magnitudes) between the stellar component and the 
stellar+disc contribution. This can be seen in Figure 
\ref{magOriC}. From this study we would expect a variability of 
$\sim$0.16 magnitudes. Kukarkin et al. (\cite{Kuk81}), in their catalogue of 
suspected variable stars, found a change in $m_{\rm v}$ of 0.06 magnitudes 
(5.10 - 5.16) for $\theta^1$ Ori C.
Hipparcos has also classified this star as variable; although Hipparcos 
data do not show a clear pattern, the median magnitude
in Hipparcos system is 4.61 mag and the variability of this magnitude
varies between 4.56 and 4.70, that is agreement with our study.
%
\subsection{Determination of the \vsini\ of $\theta^1$ Ori C}
\begin{figure}[ht!]
\centering
\includegraphics[width=6.cm,angle=90]{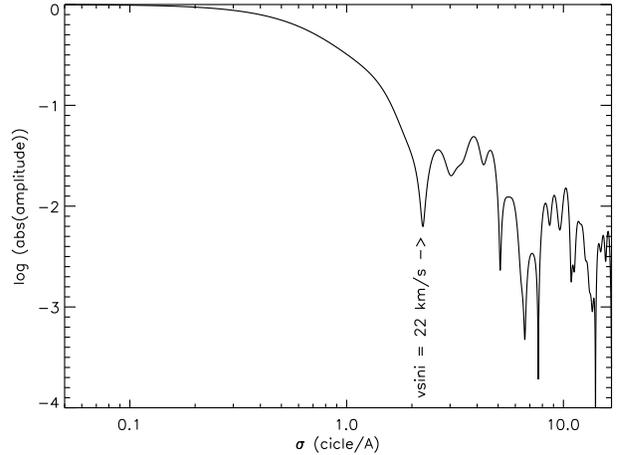}
\caption{Fourier analysis of the \ion{N}{iv}\,$\lambda$4057 line in 
         $\theta^1$ Ori C.}\label{Fourier_OriC}
\end{figure}

%
It is very important to have a good determination of the projected 
rotational velocity of this star as it is supposed that the spectral 
variability of $\theta^1$ Ori C is related with its rotation. The 
derived \vsini\ should be independent on the phase and should be 
coherent with a period of $\sim$15.4 days. It is shown in section 
\ref{variable} that the profiles of metal as well as \ion{He}{i} 
and \ion{He}{ii} lines are dependent on the phase, however this is only an 
artificial dependence due to the presence of the disk. Once the spectra 
of different phases are corrected from the effect of dilution by the disc 
continuum emission, all the metal, \ion{He}{i} and \ion{He}{ii} lines are 
independent on the phase (except those related with the disc, see 
Figure \ref{OriCvarlines}).\\
\newline
The Fourier method allows us to separate pure rotational broadening from 
other broadening mechanisms affecting the shape of the lines. We have used this 
method with some metal lines at phase $\phi$\,$\sim$\,0 (see table 
\ref{rotOriC}). 
A \vsini\,=\,24\,$\pm$\,3 \kms\ has been derived. Figure \ref{Fourier_OriC} 
shows an example of the application of the Fourier method in the 
determination of the \vsini\ of $\theta^1$ Ori C.\\
\newline
Table 7 offers a comparison of \vsini\ values obtained with the Fourier 
and \fwhm methods. We see that the Fourier method gives more consistent 
values for all considered lines. Differences within the \fwhm method may 
reach a factor of 2 and, in fact, we had problems when trying to fit 
the profile of some of the lines with a gaussian profile for measuring
their linewidth.\\ 
The derived value for \vsini\ through the Fourier method is also more 
coherent with a O7V star rotating with a 15.422 days period. 
Considering \radius\,$\sim$\,11 \radius$_\odot$, the upper limit for 
\vsini\ is $\sim$\,35 \kms and therefore the inclination
of the rotational pole is i\,$\sim$\,45, in agreement with previous
results (see next section). 
\begin{table}[h]
\begin{center}
\scriptsize{
\begin{tabular}{c c c}    
\noalign{\smallskip}
\hline
\noalign{\smallskip}
Line & \multicolumn{2}{c}{\vsini\ (\kms)} \\
     & Fourier & \fwhm \\
\noalign{\smallskip}
\hline
\noalign{\smallskip}
\ion{C}{iii} 4056 & 23 & 42 $\pm$ 10\\
\hline
\noalign{\smallskip}
\ion{N}{iii} 3998 & 23 & 28 $\pm$ 7\\
\ion{N}{iv} 4057 & 20 & 33 $\pm$ 7\\
\ion{N}{iv} 5200 & 20 & 32 $\pm$ 10\\
\hline
\noalign{\smallskip}
\ion{Si}{iv} 4089 & 23 & 62 $\pm$ 14\\
\ion{Si}{iv} 4654 & 30 & 47 $\pm$ 10\\
\hline
\noalign{\smallskip}
\ion{O}{iii} 4081 & 21 & 52 $\pm$ 10\\
\ion{O}{iii} 4376 & 26 & 31 $\pm$ 4\\
\ion{O}{iii} 4435 & 25 & 32 $\pm$ 5\\
\ion{O}{iii} 4454 & 27 & 34 $\pm$ 4\\
\ion{O}{iii} 5592 & 25 & 45 $\pm$ 6\\
\hline
\\
\end{tabular}
}
\normalsize
\rm
\caption{\footnotesize Projected rotational velocities derived 
         through Fourier and \fwhm methods for some metal lines 
		 present in the spectrum of $\theta^1$ Ori C. Values were
		 derived at phase 0.972 (see explanation in text).}
		 \label{rotOriC}
\end{center}
\end{table}
%
\subsection{Modeling of $\theta^1$ Ori C}
Once the spectral variability of $\theta^1$ Ori C is understood and
its \vsini\ has been derived, we can proceed to model its stellar 
atmosphere and wind through the observed spectrum of the star. 
The lines used for this analysis are shown in Figure \ref{HD37022_fit}; 
basically they are the ones used in the other analyses plus 
\ion{He}{i} 5875 ${\rm \AA}$. Some of the lines are contaminated 
(see Sect. \ref{variable}), so this must be taken into account. 
The H$_{\rm \delta}$ and H$_{\rm \gamma}$ lines have been selected 
as the most reliable lines for deriving \grav\ (they are less 
contaminated than H$_{\rm \beta}$ and H$_{\rm \alpha}$); The whole 
set of \ion{He}{i-ii} lines has been considered except 
\ion{He}{ii}\,$\lambda$4686; however it must be taken into account 
that the strength of all these lines vary with the phase. Phase 
0.972 will be used for determining the stellar parameters (the 
effect of the continuum emission of the disc is smaller at this 
phase, see Sect. \ref{variable}).\\
\newline
Although our study has shown that the rotational velocity 
(derived from Fourier analysis) is $\sim$\,24 \kms, when this 
broadening is considered all synthetic lines appear narrower 
than the observed ones. 
We have tried to solve the problem by means of a different 
\vsini\ value, however it does not work because then the shape of 
the synthetic profiles do not fit with the observed ones (the 
cores of the modeled profiles are too wide when the \fwhm of 
the \ion{He}{} lines is fitted). An extra-broadening mechanism 
has to be included. When a gaussian macroturbulent broadening (Gray, 
\cite{Gra73}) is used, the fit clearly improves, however in 
this case the \ion{H}{i} and \ion{He}{ii} lines cannot be fitted 
simultaneously with the \ion{He}{i} ones; when the former are 
fitted (for a \Teff\,=\,39000 K), some of the synthetic lines 
in the latter appear stronger and narrower than observed. A better
fitting for the \ion{He}{i} lines is obtained if a higher \Teff\
if considered, but then the synthetic \ion{He}{ii} lines appear too strong. 
There is no way to solve this problem in this region of the
parameter space either by changing the rotational 
velocity, the macroturbulence or the microturbulence.\\
\newline
In section \ref{individual2}, figure \ref{15Mon_fit} shows that 
the fitting of the \ion{He}{i-ii} lines could follow a similar 
behavior in the case of the O7V star HD\,47839 (selected as 
reference star). A variation of 1000 K in the effective 
temperature change strongly the strength of the \ion{He}{i}
lines. For this spectrum also occurs that when the \ion{He}{i} 
lines appear fitted, the \ion{He}{ii} lines are slightly stronger 
than observed, and if the \ion{He}{ii} lines are fitted, the
\ion{He}{i} lines are stronger than observed.\\
\newline
Puls et al. (\cite{Pul05}) have shown that there is a 
discrepancy for the \ion{He}{i} singlets between the synthetic 
\fastwind and {\sc cmfgen} lines for effective temperatures 
between 36000 and 41000 K (being the {\sc cmfgen} profiles 
shallower). Therefore the \ion{He}{i} triplet system can be 
considered more reliable (i.e. the \ion{He}{i}\,$\lambda$4471 line).\\
Knowing this discrepancy, we have considered the \ion{He}{i}\,$\lambda$4471
for the fitting with \fastwind synthetic profiles. Our best model corresponds to 
\Teff\,=\,39000\,$\pm$\,1000 K and \grav\,=\,4.1 dex. From these values, 
the model flux and with $M_{\rm v}$\,=\,-4.9\,$\pm$\,0.3, a stellar radius 
\radius\,=\,10.6\,$\pm$\,1.5 \radius$_{\sun}$ is derived, and then an 
inclination of the rotational axis of $i$\,=\,44\,$\pm$\,12$^{\circ}$. 
This value is in agreement with previous independent studies 
(Stahl et al. \cite{Sta96}, Donati et al. \cite{Don02}), although 
our derived \vsini\ is more reliable and our radius is not obtained 
from a SpT - \radius\ calibration but is a result of the spectral 
analysis of the star.
\begin{figure*}[ht!]
\centering
\includegraphics[width=12.cm,angle=90]{./HD37022_p0972_fit.e
ps}
\caption{HHe analysis of HD\,37022 ($\theta^1$ Ori C, O7Vp) at phase $\phi$\,=\,0.972. 
         The solid black line corresponds to a model with \Teff,=\,39000 K, \grav\,=\,4.1
		 $\epsilon$\,=\,0.09, $\xi_{\rm t}$\,=\,5 \kms\ and log\,$Q$\,=\,-13.0;
		 the dashed blue line corresponds to a model with \Teff\,=40000 K (same remaining
		 parameters). A \vsini\,=\,24 \kms\ and a macroturbulent velocity of 60 \kms\
		 have been considered. The observed lines strength has been enlarged by a 
		 factor 1.1 to correct for the effect of the disc continuum emission (see text
		 for explanation).}
\label{HD37022_fit}
\end{figure*}
\section{Oxygen abundances in the Trapezium cluster B0.5V stars}
\label{abundance}
%
Three of the stars studied in Orion are perfect targets for a stellar 
abundance analysis as they have many narrow unblended lines. 
These are HD\,37020, HD\,37023 and HD\,37042, three B0.5V stars. Cunha
\& Lambert (\cite{Cun92}, \cite{Cun94}) presented carbon, nitrogen, 
oxygen and silicon abundances from {\sc LTE} and {\sc NLTE} analyses 
for these stars. For comparison purposes, a similar analysis has been 
done for 
$\tau$ Sco, a B0.2V star (Walborn \& Fitzpatrick, \cite{Wal90}) with very 
low \vsini. Stellar abundances for
this star have been derived elsewhere in the literature (Hardorp \& Scholz 
\cite{Har70}, Kane et al. \cite{Kan80}, Peters \& Polidan \cite{Pet85},  
Sch\"onberner et al. \cite{Sch88}, Becker \& Butler \cite{Bec88}, Kilian et
al. \cite{Kil91}, Martin \cite{Mar04}, see Table \ref{t_comp_tau}). 
The other two Orion stars have been ruled 
out: HD\,37041 has a relatively high projected rotational velocity, so 
metallic lines are broadened and then appear blended; HD\,37022, being 
an O7V star, does not have enough oxygen lines for an accurate abundance
analysis.\\
\newline
We have therefore derived oxygen abundances for the Trapezium cluster B0.5V stars 
for comparing them with the \orion\ nebular abundances obtained by Esteban 
et al. (\cite{Est04}).
%
\subsection{Line identification and measurement of equivalent widths}
\label{EW}
We make use of the classical method of curve of growth to determine 
the oxygen abundances. When this methodology is used it is important 
to remove all lines that appear blended. The
spectrum of $\tau$ Sco (a star with similar spectral type than our 
targets in Orion and a very low \vsini) has been used for an
identification of the \ion{O}{ii} lines present in the spectra of the 
Orion stars. The whole set of lines is shown in Table \ref{LinesO}
divided into multiplets, together with them log\,$gf$ values (basically 
they are taken from the NIST database).\\
The equivalent widths of all the \ion{O}{ii} lines listed in Table 
\ref{LinesO} have been measured for the four stars, however only our set 
of suitable lines (see below) is shown in Table \ref{t_O_EW}. To measure 
the equivalent widths we use our own software developed in IDL. A least 
squares profile fitting procedure was used, with Gaussian profiles 
fitting the line and polynomials of degree one or two to fit the local 
continuum. Errors in the measurements due to the uncertainty in the 
position of the local continuum (estimated as $\pm$1/\snr, Villamariz 
et al. \cite{Vil02}) have also been considered. The estimated value of 
the uncertainty in the measurement of the \ew s is $\sim$5 m{\rm \AA}
and $\sim$10 m{\rm \AA} for some problematic lines.
%
\subsection{Line selection}\label{selection}
%
\begin{table*}[t!]
\begin{center}
\scriptsize{
\begin{tabular}{c r c l r c l r c l r c l}    
\hline
\noalign{\smallskip}
Line & \ew$_{\rm o}$ & $\epsilon$(O)$^a$ & $\Delta\epsilon$(O) &
     \ew$_{\rm o}$ & $\epsilon$(O) & $\Delta\epsilon$(O) &
     \ew$_{\rm o}$ & $\epsilon$(O) & $\Delta\epsilon$(O) &
     \ew$_{\rm o}$ & $\epsilon$(O) & $\Delta\epsilon$(O) \\
\noalign{\smallskip}
\hline
\hline
\noalign{\smallskip}
\noalign{\smallskip}
 & \multicolumn{3}{c}{HD\,149438 ($\xi_t$ = 8.7)} 
 & \multicolumn{3}{c}{HD\,37020  ($\xi_t$ = 6.5)} 
 & \multicolumn{3}{c}{HD\,37023  ($\xi_t$ = 7.4)} 
 & \multicolumn{3}{c}{HD\,37042  ($\xi_t$ = 5.5)} \\
\noalign{\smallskip}
\cline{2-13}
\noalign{\smallskip}
\noalign{\smallskip}
 & \multicolumn{3}{c}{\Teff\,=\,32000 K \,\,\, \grav\,=\,4.0} &
  \multicolumn{3}{c}{\Teff\,=\,30000 K \,\,\, \grav\,=\,4.0} &
  \multicolumn{3}{c}{\Teff\,=\,32000 K \,\,\, \grav\,=\,4.2} &
  \multicolumn{3}{c}{\Teff\,=\,29000 K \,\,\, \grav\,=\,4.2} \\
\noalign{\smallskip}
\cline{1-13}
\noalign{\smallskip}
\noalign{\smallskip}
\ion{O}{ii}\,4638 &  85 & 8.75 & 0.04 
                             &  --- & --- & ---
  						     &  --- & --- & ---
						     &  --- & --- & ---\\
\ion{O}{ii}\,4641 & 127 & 8.65 & 0.04
                             & --- & --- & ---
							 & --- & --- & ---
							 & --- & --- & --- \\
\ion{O}{ii}\,4661 &  90 & 8.70 & 0.04
                             &  80 & --- & ---
							 &  76 & 8.60 & 0.06
							 &  88 & 8.56 & 0.06 \\
\ion{O}{ii}\,4676 &  79 & 8.71 & 0.05
                             &  82 & 8.59 & 0.06
							 &  65 & 8.59 & 0.06
							 &  82 & 8.63 & 0.06 \\
\ion{O}{ii}\,4317 &  92 & 8.69 & 0.03
                             & 107 & 8.67 & 0.04
							 &  78 & 8.59 & 0.05
							 &  98 & 8.60 & 0.07 \\
\ion{O}{ii}\,4319 &  86 & 8.71 & 0.04
                             & 100 & 8.69 & 0.05
							 &  75 & 8.63 & 0.05
							 &  95 & 8.66 & 0.07 \\
\ion{O}{ii}\,4366 &  80 & 8.61 & 0.04
                             &  96 & 8.62 & 0.05
							 &  77 & 8.62 & 0.05
							 & 103 & 8.62 & 0.08 \\
\ion{O}{ii}\,4416 & 100 & 8.62 & 0.04
                             & 120 & 8.65 & 0.05
							 &  93 & 8.62 & 0.05
							 & 117 & 8.70 & 0.11 \\
\ion{O}{ii}\,4452 &  37 & 8.67 & 0.07
                             &  48 & 8.64 & 0.07
							 &  32 & 8.58 & 0.09
							 &  60 & 8.67 & 0.10 \\
\ion{O}{ii}\,4072 &  --- & --- & ---
                             & 118 & 8.61 & 0.12
							 &  99 & 8.57 & 0.09
							 & 119 & 8.71 & 0.10 \\
\ion{O}{ii}\,4076 &  --- & --- & ---
                             & 133 & 8.59 & 0.12
							 & 114 & 8.57 & 0.09
							 & --- & --- & --- \\
\ion{O}{ii}\,4078 &  --- & --- & ---
                             &  46 & 8.67 & 0.17
							 &  33 & 8.59 & 0.09
							 &  48 & 8.68 & 0.10 \\							 
\ion{O}{ii}\,4941 &  38 & 8.64 & 0.06
                             &  50 & 8.68 & 0.07
							 &  35 & 8.60 & 0.09
							 &  43 & 8.60 & 0.08 \\
\ion{O}{ii}\,4943 &  60 & 8.62 & 0.04
                             &  67 & 8.58 & 0.06
							 &  55 & 8.59 & 0.07
							 &  56 & 8.48 & 0.07 \\
\ion{O}{ii}\,4956 &  17 & 8.73 & 0.12
                             &  20 & 8.67 & 0.14
							 &  13 & 8.58 & 0.20
							 &  22 & 8.71 & 0.13 \\
\ion{O}{ii}\,4891 &  24 & 8.75 & 0.10
                             &  27 & 8.68 & 0.11
							 &  18 & 8.59 & 0.15
							 &  27 & 8.65 & 0.12 \\
\ion{O}{ii}\,4906 &  34 & 8.64 & 0.07
                             &  42 & 8.63 & 0.08
							 &  31 & 8.59 & 0.10
							 &  47 & 8.72 & 0.08\\
\noalign{\smallskip}
\hline
\noalign{\smallskip}
 & \multicolumn{3}{c}{$\epsilon$(O) = 8.70 $\pm$ 0.10} &
   \multicolumn{3}{c}{$\epsilon$(O) = 8.65 $\pm$ 0.10} &
   \multicolumn{3}{c}{$\epsilon$(O) = 8.59 $\pm$ 0.10} &
   \multicolumn{3}{c}{$\epsilon$(O) = 8.64 $\pm$ 0.10} \\
\noalign{\smallskip}
\cline{2-13}
\\
\multicolumn{3}{l}{$^a$ $\epsilon$(O) = log(O/H) + 12} & & & & & & & & & & \\
\\
\end{tabular}
}
\normalsize
\rm
\caption{\footnotesize Equivalent widths and derived line abundances
                       for the set of \ion{O}{ii} lines used in our analysis. 
					   Line abundances refer to
					   the microturbulence given in brackets for each 
					   star ($\xi_t$ in \kms). Uncertainties in the
					   line abundances come from the propagation of 
					   the uncertainties of the equivalent width 
					   measurements (see text). Some of the \ion{O}{ii}
					   lines of the Orion stars have not been used as 
					   they appear blended.\ion{O}{ii}\,$\lambda\lambda$
					   4072, 4076 and 4078 lines were ruled out in the analysis of
					   $\tau$ Sco due to the poor quality of the {\sc caspec} spectrum
					   in this region. Final oxygen abundances for each star have 
					   been calculated through a weighted mean of the linear values. 
					   Errors represent the statistical deviation for these mean values.
\label{t_O_EW}
}
\end{center}
\end{table*}
Some of the lines that appear unblended in the spectrum of $\tau$ Sco 
cannot be used in the analyses of the other stars; they 
have larger rotational broadening and then appear blended (or lie in the 
wings of \ion{H}{} or \ion{He}{} lines). Once a first set of unblended 
lines was selected, a preliminary abundance analysis was done separately 
for each multiplet. In this way the dispersion in the line abundances for
the zero slope value of the microturbulence are minimized as all the 
lines in a multiplet are formed in the same region in the stellar 
atmosphere. Problematic lines, errors in the measurement of the equivalent 
widths or artificial trends can be then identified; such lines will be 
removed in the global analysis (e.g. this is the case for the 
\ion{O}{ii}\,$\lambda$4414 line, an isolated line that gives systematically 
lower abundances).\\
\newline
The set of suitable lines finally used in the abundance analyses is 
presented in Table \ref{t_O_EW}. They are lines coming from 
transitions between configurations 2p$^2$\,($^3$P)\,3s\,-\,2p$^2$\,($^3$P)\,3p 
(NIST multiplets 64, 65 and 72) and 2p$^2$\,($^3$P)\,3p\,-\,2p$^2$\,($^3$P)\,3d 
(NIST multiplets 90, 148 and 130). 
We have ruled out the lines that do not follow the general trend. We
have found that lines from multiplets 99, 118, 161, 188, 172 give
systematically lower abundances. This effect can be due to
the definition of the oxygen model atom we have used, or can be associated
with the log\,$gf$ values. Some preliminary comparisons with {\sc tlusty} 
analyses have shown that the difference in the line abundances
is also present for the case of $\tau$ Sco.
\subsection{Chemical analysis}\label{chemical}
\begin{figure}[ht!]
\includegraphics[width=6.cm,angle=90]{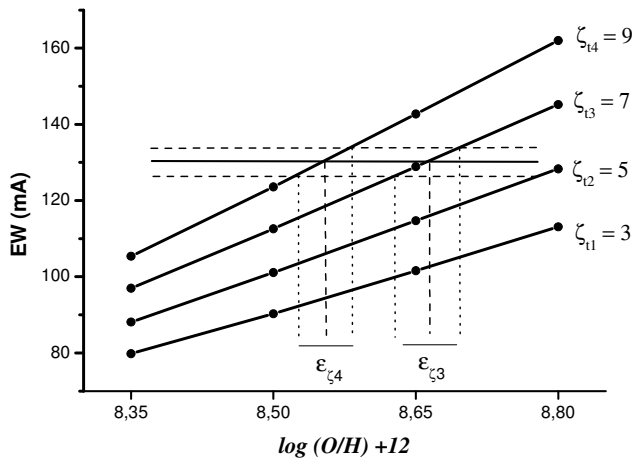}
\caption{Example of curve of grow for the line \ion{O}{ii}\,$\lambda$4414
         in the star HD\,37042. A grid of 16 models has been considered
		 (4 microturbulences and 4 oxygen abundances). The observed \ew\ of
		 the line and its uncertainty are plotted as horizontal lines. 
		 Two examples of abundances (and their uncertainties) derived for 
		 microturbulences 7 and 9 \kms\ are plotted as vertical lines.}
\label{OII4414_method1}
\end{figure}

\begin{figure}[ht!]
\includegraphics[width=6.5cm,angle=90]{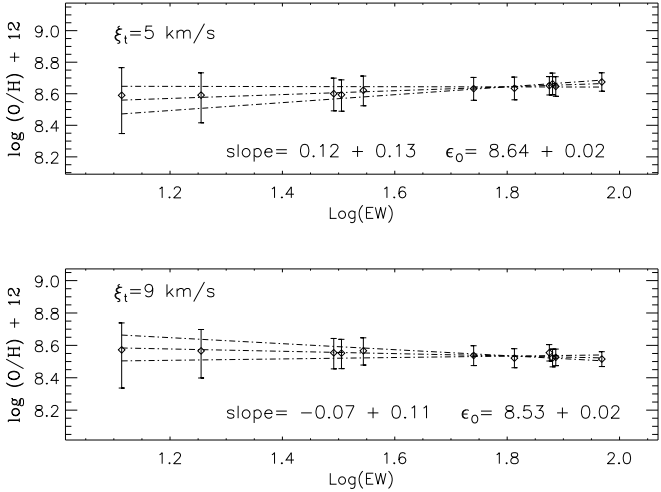}
\caption{Example of log\,$\epsilon$\,-\,\ew\ diagrams in the study of 
         HD\,37023.}
\label{curveog2}
\end{figure}

For each star we proceed as follows. A grid of 16 \fastwind models 
combining four abundances and four values of microturbulence is 
calculated. In this way, the curves of growth for each line can be 
constructed by plotting the theoretical equivalent width for each 
abundance and microturbulence versus the abundance (see Figure
\ref{OII4414_method1} for the case of the \ion{O}{ii}\,$\lambda$\,4414 
line in HD\,37042).\\ 
Through the observed equivalent width and its error one abundance 
(and its derived uncertainty) can be derived for each line and 
microturbulence. The individual line abundances are dependent on the 
microturbulence. The microturbulence value that minimizes the 
dependence of the line abundances on the line strength in the 
log\,$\epsilon$\,-\,\ew\ diagrams will be the microturbulence we 
are looking for. These diagrams can be also used as a diagnostic tool 
to check the reliability of the different lines for the abundance 
determination (see Section \ref{selection}). 
Figure \ref{curveog2} shows the log\,$\epsilon$\,-\,\ew\ diagrams for 
two different microturbulences in the study of HD\,37023.\\
\newline
The microturbulence derived from the zero slope for each star is presented
in Table \ref{Oabun}. Uncertainties in the microturbulence are obtained
considering the errors derived for the slope in a linear fit of the data 
(due to errors in the individual line abundances). This step also allows 
us to estimate the contribution of the uncertainty in the microturbulence 
to the the total oxygen abundance. This uncertainty depends mainly
on the quality of the spectra (in this case is $\sim$ 0.06 dex).\\
\newline
\newline
Abundance values for each line as well as their uncertainties are obtained 
for that microturbulence (see Table \ref{t_O_EW}). The final abundance 
value is calculated through a weighted mean of the linear individual line 
abundances (10$^{\,\epsilon_i-12}$), and its uncertainty ($\sigma_{\rm w}$) 
is that associated with this mean. 
Final values for the total oxygen abundances for each star are shown 
in Table \ref{Oabun}. The final uncertainty in the oxygen abundance 
takes into account four different sources of errors: those associated 
with the statistical analysis, those derived from the error in the 
determined microturbulence and finally those refered to the uncertainty 
in the stellar parameters and to atomic data. All these sources of 
error are added quadratically for deriving the final abundance 
uncertainty (see Villamariz et al. \cite{Vil02}).\\
\begin{table}[ht!]
\begin{center}
\scriptsize{
\begin{tabular}{c c c c}    
\noalign{\smallskip}
\hline
\noalign{\smallskip}
Object & N & $\xi_{\rm t}$ (km/s) & log(O/H)+12\\
\hline
\noalign{\smallskip}
 & & & This work\\
\hline
\noalign{\smallskip}
 HD\,149438 & 14 & 8.1 $\pm$ 1.2 & 8.70 $\pm$ 0.10 \\
\noalign{\smallskip}
\noalign{\smallskip}
 HD\,37020  & 13 & 6.8 $\pm$ 1.3 & 8.65 $\pm$ 0.10 \\
 HD\,37023  & 15 & 6.3 $\pm$ 3.0 & 8.59 $\pm$ 0.10 \\
 HD\,37042  & 13 & 4.9 $\pm$ 2.0 & 8.64 $\pm$ 0.10 \\
\noalign{\smallskip}
\noalign{\smallskip}
\hline
\noalign{\smallskip}
 & & \multicolumn{2}{c}{Cunha $\&$ Lambert ({\sc LTE})}\\
\hline
\noalign{\smallskip}
 HD\,37020  & 6 & 7.0 & 8.92 $\pm$ 0.05\\
 HD\,37023  & 7 & 7.5 & 8.76 $\pm$ 0.04\\
 HD\,37042  & 7 & 6.0 & 8.97 $\pm$ 0.04\\
\noalign{\smallskip}
\noalign{\smallskip}
\hline
\noalign{\smallskip}
 & & \multicolumn{2}{c}{Cunha $\&$ Lambert ({\sc NLTE})}\\
\hline
\noalign{\smallskip}
 HD\,37020  & 7 & 5.0 & 8.83 $\pm$ 0.12\\
 HD\,37023  & 6 & 7.0 & 8.87 $\pm$ 0.08\\
 HD\,37042  & 6 & 6.0 & 8.85 $\pm$ 0.06\\
\noalign{\smallskip}
\noalign{\smallskip}
\hline
\noalign{\smallskip}
 & & & Esteban et al.\\
\hline
\noalign{\smallskip}
Gas & & & 8.65 $\pm$ 0.03 \\
Gas + dust & & & 8.73 $\pm$ 0.03\\
\hline
\\
\end{tabular}
}
\normalsize
\rm
\caption{\footnotesize Oxygen abundances for the three B0.5V stars 
inside Orion Nebula and the reference star $\tau$ Sco. Oxygen 
{\sc LTE} and {\sc NLTE} abundances derived by Cunha $\&$ Lambert 
(\cite{Cun94}) for the Orion stars as well as those calculated by 
Esteban et al. (\cite{Est04}) for the nebula are also presented 
for comparison.
\label{Oabun}
}
\end{center}
\end{table}
%
\subsection{Oxygen abundance in the stardard star $\tau$ Sco}
\label{tausco}
Table \ref{t_comp_tau} sumarizes the oxygen abundances appeared
in the literature for $\tau$ Sco. Our derived value is 
compatible with previous results but a little higher
than most of them. This difference can be easily explained
taking into account that for this range of stellar parameters,
the oxygen abundance derived through \ion{O}{ii} lines is very 
sensitive to a change in \Teff\ and \grav\ (the lines become fainter when 
a higher \Teff\ is considered and then the derived oxygen 
abundance is higher). The uncertainty in the oxygen abundance 
due to a change of $\sim$\,1000 K in \Teff\ is $\sim$\,0.08 dex. 
This effect is considered in the uncertainty for the given values,
however the central value will be slightly dependent on the 
derived stellar parameters.
We have considered two \fastwind models with different \Teff\ (32000,
32500 K) and the same \grav\, obtaining oxygen abundances of
8.70 and 8.74 dex respectively.\\
\\
The derived abundance is also very dependent on the microturbulence
considered (specially if lines with high equivalent width are 
used). We have taken into account these dependencies in our 
uncertainties.
\begin{table}[!ht]
\begin{center}
\scriptsize{
\begin{tabular}{c c c c c}    
\noalign{\smallskip}
\hline
\noalign{\smallskip}
 Work & \Teff\ (K) & \grav\ & $\epsilon$(O) & $\xi_{\rm t}$ (\kms)\\
\noalign{\smallskip}
\hline
\noalign{\smallskip}
Hardorp \& Scholz & 30900 & 4.05 & 8.70 & --- \\
Kane et al.  & 30300 & 3.95 & 8.6 & --- \\
Peters \& Polidan & 31500 & 4.3 & 8.65\,$\pm$\,0.26 & 5 \\
Sch\"onberner et al. & 33000 & 4.15 & 8.5 & 5 \\
Becker \& Butler & 33000 & 4.15 & 8.70\,$\pm$\,0.19 & 5 \\
                 &       &      & 8.58\,$\pm$\,0.19 & 10 \\
Kilian et al. & 31400 & 4.24 & 8.61\,$\pm$\,0.12 & 3 \\
Martin & 30000 & 3.75 & 8.58\,$\pm$\,0.17 & 7 \\
This work & 32000 & 4.0 & 8.70\,$\pm$\,0.10 & 8 \\
\noalign{\smallskip} 
\hline
\\
\end{tabular}
}
\normalsize
\rm
\caption{\footnotesize Comparison of stellar parameters and 
         abundances derived for $\tau$ Sco in previous studies
         found in the literature and in this work.
\label{t_comp_tau}
}
\end{center}
\end{table}
%
\subsection{Comparing with nebular and previous stellar results}
\label{nebular}
The oxygen abundances derived for the Orion stars are 
compatible within the errors (see Table \ref{Oabun}). HD\,37023
has a slightly lower abundance (but still compatible with
the other abundances). In section \ref{individual} we have seen that
the fitting of the H and He lines is not so good as for the
other B0.5V stars (observed lines appear slightly broader).\\
The oxygen abundances in the Orion stars are systematically
lower than that derived for $\tau$ Sco.\\ 
\newline 
Esteban et al. (\cite{Est04}) have recently published a reappraisal 
of the chemical composition of the Orion nebula. They derived a 
total oxygen gas-phase abundance $\epsilon$(O)\,=\,8.65\,$\pm$\,0.03. 
However, some oxygen is expected to be depleted onto dust grains in
ionized nebula, so the total 
gas+dust oxygen abundance should take this depletion into account. 
In a previous work (Esteban et al. \cite{Est98}), these authors 
estimate the depletion onto dust grains by comparing Si and Fe nebular 
abundances with those obtained by Cunha $\&$ Lambert (\cite{Cun94}) 
for B stars in the Orion association, assuming a certain composition
for the main dust molecules. Taking 
into account this correction, the final gas+dust oxygen abundance 
Esteban et al. (\cite{Est04}) propose is 
$\epsilon$(O)\,=\,8.73\,$\pm$\,0.03 (an oxygen abundance correction for 
dust $\sim$\,0.08 is applied).\\
\newline
Our stellar results are compatible with those obtained by Esteban 
et al. (\cite{Est04}) for the gas phase, however the dust+gas 
corrected abundance seems to be too high compared with our derived 
stellar abundances, although still marginally consistent inside the 
uncertainties.\\
\newline
Our oxygen abundances are systematically lower than the NLTE abundances
by Cunha $\&$ Lambert (\cite{Cun94}). In that paper the former authors
comment that the LTE abundances are slightly more reliable than
the NLTE abundances they present. The difference between their LTE 
abundances and our results are even higher. These differences can be in
part associated with the differences in the derived stellar parameters
for these stars. The \Teff\ and \grav\ obtained by these authors are 
higher than ours (see Table \ref{Cunhatable}), so the oxygen abundances
they derive are obviously higher.\\
This difference in the effective temperatures may also affect the
derived silicon abundances (used by Esteban et al. \cite{Est98} for
estimating the oxygen depletion). Cunha \& Lambert (\cite{Cun94}), derive
their silicon abundances by using 3 \ion{Si}{iii} lines. Preliminary
silicon analysis by our group has shown that a difference of $\sim$\,1000 K
in \Teff\ can shift the \ion{Si}{iii} abundances up to 0.2 dex (deriving a
higher abundance for the lower \Teff).\\
\newline
Alternatively, our result could suggest that the molecules that Esteban et al.
used to estimate the O dust depletion in Orion cannot be present in 
large amounts in this nebula. Consequently, Si, Mg and Fe (the refractory
elements being the main constituents of those molecules) have to form
other molecules without oxygen.\\
\begin{table}[ht!]
\begin{center}
\scriptsize{
\begin{tabular}{c c r c}    
\noalign{\smallskip}
\hline
\noalign{\smallskip}
Label & $\lambda$ ({\rm \AA}) & log\,{\em gf} & Notes \\
\noalign{\smallskip}
\hline
\noalign{\smallskip}
\multicolumn{4}{l}{2s$^2$ 2p$^2$ (3P) 3p - 2s$^2$ 2p$^2$ (3P) 3d} \\
\noalign{\smallskip}
\hline
\noalign{\smallskip}
 & & \multicolumn{2}{r}{NIST 90 (4D$_0$-4F)}\\
\noalign{\smallskip}
\hline
\noalign{\smallskip}
\ion{O}{ii}\,4069 & 4069.623 & 0.149  & + \ion{C}{iii} \\
                  & 4069.882 & 0.344  &   \\
\ion{O}{ii}\,4072 & 4072.153 & 0.552  &   \\
\ion{O}{ii}\,4076 & 4075.862 & 0.693  & + \ion{C}{ii}\ \\
\ion{O}{ii}\,4078 & 4078.842 & -0.284 &   \\
\noalign{\smallskip}
\hline
\noalign{\smallskip}
 & & \multicolumn{2}{r}{NIST 118 (2D$_0$-2F)} \\
\noalign{\smallskip}
\hline
\noalign{\smallskip}
\ion{O}{ii}\,4699 & 4699.218  & 0.269  &  \\
                  & (4699.011 & 0.418) &  NIST 172\\
                  & (4698.437 & -0.883) & NIST 172 \\
\ion{O}{ii}\,4705 & 4705.346  & 0.477  &  \\
\ion{O}{ii}\,4741 & 4741.704  & -0.987 & Very weak  \\
\noalign{\smallskip}
\hline
\noalign{\smallskip}
 & & \multicolumn{2}{r}{NIST 130 (4S$_0$-4P)} \\
\noalign{\smallskip}
\hline
\noalign{\smallskip}
\ion{O}{ii}\,4891 & 4890.856 & -0.436 &  \\
\ion{O}{ii}\,4906 & 4906.830 &-0.160  &  \\
\noalign{\smallskip}
\hline
\noalign{\smallskip}
 & & \multicolumn{2}{r}{NIST 148 (2P$_0$-2D)} \\
\noalign{\smallskip}
\hline
\noalign{\smallskip}
\ion{O}{ii}\,4941 & 4941.072 & -0.054 &  \\
\ion{O}{ii}\,4943 & 4943.005 & 0.239  &  \\
\ion{O}{ii}\,4956 & 4955.707 & -0.573 & Very weak \\
\noalign{\smallskip}
\hline
\noalign{\smallskip}
\multicolumn{4}{l}{2s$^2$ 2p$^2$ (3P) 3s - 2s$^2$ 2p$^2$ (3P) 3p}\\
\noalign{\smallskip}
\hline
\noalign{\smallskip}
 & & \multicolumn{2}{r}{NIST 64 (4P-4D$_0$)} \\
\noalign{\smallskip}
\hline
\noalign{\smallskip}
\ion{O}{ii}\,4638 & 4638.856 & -0.332 & + \ion{C}{ii} + \ion{Si}{iii}\\
\ion{O}{ii}\,4641 & 4641.810 & 0.054  & + \ion{N}{iii}\\
\ion{O}{ii}\,4650 & 4649.135 & 0.308  & \\
\ion{O}{ii}\,4651 & 4650.638 & -0.362 & \\
\ion{O}{ii}\,4661 & 4661.632 & -0.278 & \\
\ion{O}{ii}\,4673 & 4673.733 & -1.088 & \\
\ion{O}{ii}\,4676 & 4676.235 & -0.394 & \\
\ion{O}{ii}\,4696 & 4696.352 & -1.380 & \\
\noalign{\smallskip}
\hline
\noalign{\smallskip}
 & & \multicolumn{2}{r}{NIST 65 (4P-4P$_0$)}\\
\noalign{\smallskip}
\hline
\noalign{\smallskip}
\ion{O}{ii}\,4317 & 4317.139 & -0.386 &  \\
                  & 4317.696 & ...    &  \\
\ion{O}{ii}\,4319 & 4319.630 & -0.380 &  + \ion{N}{iii}\\
                  & 4319.866 & -0.502 &  \\
\ion{O}{ii}\,4366 & 4366.895 & -0.348 &  \\
                  & 4366.530 & ...    &  \\
\noalign{\smallskip}
\hline
\noalign{\smallskip}
 & & \multicolumn{2}{r}{NIST 72 (2P-2D$_0$)}\\
\noalign{\smallskip}
\hline
\noalign{\smallskip}
\ion{O}{ii}\,4414 & 4414.899 & 0.172  &  \\
                  & 4414.456 & -1.483 &  \\
\ion{O}{ii}\,4416 & 4416.975 & -0.076 & + \ion{Si}{iv} \\
\ion{O}{ii}\,4452 & 4452.378 & -0.789 &  \\
\noalign{\smallskip}
\hline
\\
\end{tabular}
}
\normalsize
\rm
\caption{\footnotesize Preliminary set of \ion{O}{ii} lines selected 
         for the analysis, divided by multiplets. The spectrum 
		 of the low \vsini\ star $\tau$ Sco, has been used 
         to identify the lines. log\,{\em gf} values are from the NIST 
		 database.}\label{LinesO}
\end{center}
\end{table}

\begin{table}[ht!]
\begin{center}
\scriptsize{
\begin{tabular}{c c r c}    
\noalign{\smallskip}
\hline
\noalign{\smallskip}
Label & $\lambda$ ({\rm \AA}) & log\,{\em gf} & Notes \\
\noalign{\smallskip}
\hline
\noalign{\smallskip}
\multicolumn{4}{l}{2s$^2$ 2p$^2$ (1D) 3s - 2s$^2$ 2p$^2$ (1D) 3p}\\
\noalign{\smallskip}
\hline
\noalign{\smallskip}
 & & \multicolumn{2}{r}{NIST 99 (2D-2F$_0$)} \\
\noalign{\smallskip}
\hline
\noalign{\smallskip}
\ion{O}{ii}\,4590 & 4590.974 & 0.350  &  \\
\ion{O}{ii}\,4596 & 4595.957 & -1.032 &  \\
                  & 4596.177 & 0.200  &  \\
\noalign{\smallskip}
\hline
\noalign{\smallskip}
\multicolumn{4}{l}{2s$^2$ 2p$^2$ (1D) 3p - 2s$^2$ 2p$^2$ (1D) 3d}\\
\noalign{\smallskip}
\hline
\noalign{\smallskip}
 & & \multicolumn{2}{r}{NIST 161 (2F$_0$-2G)} \\
\noalign{\smallskip}
\hline
\noalign{\smallskip}
\ion{O}{ii}\,4185 & 4185.440 & 0.604  & + \ion{C}{iii} \\
\ion{O}{ii}\,4189 & 4189.788 & 0.716  &  \\
                  & 4189.581 & -0.828 &  \\
\hline
\noalign{\smallskip}
 & & \multicolumn{2}{r}{NIST 188 (2P$_0$-2P)} \\
\noalign{\smallskip}
\hline
\noalign{\smallskip}
\ion{O}{ii}\,4691 & 4690.888 & -0.610 &  \\
                  & 4691.419 & -0.309 &  \\
\ion{O}{ii}\,4701 & 4701.179 & 0.088  &  \\
                  & 4701.712 & -0.611 &  \\
                  & 4700.441 & -3.298 &  \\
\hline
\noalign{\smallskip}
 & & \multicolumn{2}{r}{NIST 172 (2D$_0$-2F)}\\
\noalign{\smallskip}
\hline
\noalign{\smallskip}
\ion{O}{ii}\,4703 & 4703.161 & 0.263 &  \\
\hline
\\
\end{tabular}
}
\normalsize
\rm
\caption{\footnotesize (Cont) Preliminary set of \ion{O}{ii} lines selected 
         for the analysis, divided by multiplets. The spectrum 
		 of the low \vsini\ star $\tau$ Sco, has been used 
         to identify the lines. log\,{\em gf} values are from the NIST 
		 database.}\label{LinesO}
\end{center}
\end{table}
%
\section{Conclusions}\label{conclude}
By means of a detailed spectroscopic analysis of the optical
spectra of the Trapezium cluster stars, stellar parameters
and oxygen abundances have been derived. Projected rotational 
velocities has been obtained through Fourier method. This
method has been extensively used for late type stars, but
not widely applied to early type stars. Our results show
this method to be very useful for distinguishing between rotational
broadening and another broadening mechanisms that can be present in
early type stars (i.e. macroturbulence). The agreement is very
good when comparing with results from the line-width method.
The Fourier method applied to the high resolution $\theta^1$ Ori C 
\feros\ spectra,
allow us to derive a very accurate \vsini\ that is in agreement
with the period of variability of some spectral features
in $\theta^1$ Ori C.\\
Stellar parameters and their uncertainties have been derived
for the studied stars using \ion{H}{}, \ion{He}{i} and \ion{He}{ii}
lines and \fastwind\ code.\\
\newline
The presence of many \ion{O}{ii} lines in the optical 
spectrum of three B0.5V Orion stars has allowed us to work
on a very detailed abundance analysis using the curve of
growth method. This analysis has been performed through a 
careful selection of suitable lines from a previous
study of the different \ion{O}{ii} multiplets. In this way,
the dispersion in the line abundances is reduced, and the 
final abundance value derived is very precise.\\
\newline
The derived oxygen abundances in the Orion stars are in
agreement with the nebular gas-phase abundances obtained
by Esteban et al. (\cite{Est04}), and $\sim$\,0.2 dex
lower than the NLTE abundances derived by Cunha \& Lambert 
(\cite{Cun94}).\\
The gas+dust corrected oxygen abundances estimated by
Esteban et al. (\cite{Est98}, \cite{Est04}) using the 
Cunha \& Lambert stellar abundances in the Orion 
association, seem to be too high compared with our
derived abundances (although still marginally consistent 
within the uncertainties). This result suggests a lower
dust depletion factor of oxygen than previous estimations
for the Orion nebula. A revision of  the silicon, magnesium 
and iron stellar abundances in the Trapezium cluster stars 
is presently under way in our group for confirming
this result.\\ 

\begin{acknowledgements}
We want to thank M. A. Urbaneja for the original procedures to derive 
the stellar abundances and his invaluable help with \fastwind; D. Lennon
for his comments and also P. Dufton and R. Ryans for calculating some
{\sc TLUSTY} models for comparison. This work has been partially funded
by the Spanish Ministerio de Educaci\'on y Ciencia under projects 
AYA2004-08271-C02-01 and AYA2004-07466.
We are very grateful to T. Gehren and O. Stahl for lend us the spectra
of $\tau$ Sco and $\theta^1$ Ori C. This research has made use of the
{\sc eso-feros} database.

\end{acknowledgements}




\begin{thebibliography}{}
 \bibitem[1997]{Bab97}
 Babel, J., \& Montmerle, T. 1997, ApJ, 485, L29
 \bibitem[1988]{Bec88}
 Becker, S.R., \& Butler K. 1988, A\&A, 201, 232
 \bibitem[1994]{Cai94}
 Caillault, J.P., Gagne, M., \& Stauffer J.R. 1994, ApJ, 432, 386
 \bibitem[1933]{Car33}
 Carroll, J.A. 1933, MNRAS, 93, 478
 \bibitem[1972]{Con72}
 Conti, P.S. 1972, ApJ, 174, L79
 \bibitem[2002]{Cro02}
 Crowther, P.A., Hillier, D.J., Evans, C.J., et al. 2002, ApJ, 579, 774
 \bibitem[1992]{Cun92}
 Cunha, K., \& Lambert, D.L. 1992, ApJ, 399, 586
 \bibitem[1994]{Cun94}
 Cunha, K., \& Lambert, D.L. 1994, ApJ, 426, 170
 \bibitem[2002]{Don02}
 Donati, J.F., Babel, J., Harries, T.J., et al. 2002, MNRAS, 333, 55
 \bibitem[1998]{Est98}
 Esteban, C., Peimbert, M., Torres-Peimbert, S., et al. 1998, MNRAS, 295, 401
 \bibitem[2004]{Est04}
 Esteban, C., Peimbert, M., Garc\'{\i}a-Rojas, J., et al. 2004, MNRAS, 355, 229
 \bibitem[2001]{Fer01}
 Ferland, G.F. 2001, PASP, 113, 165
 \bibitem[1997]{Gag97}
 Gagn\'e, M., Caillault, J.P., Stauffer, J.R., et al. 1997, ApJ, 478, L87
 \bibitem[1993]{Gie93}
 Gies, D.R., Mason, B.D., Hartkopf, W.I., et al. 1993, AJ, 106, 2072
 \bibitem[1973]{Gra73}
 Gray, D.F. 1973, ApJ, 184, 461
 \bibitem[1971]{Har70}
 Hardorp, J., Scholz, M. 1970, ApJS, 19, 193
 \bibitem[1992]{Her92}
 Herrero, A., Kudritzki, R. P., Vilchez, J. M., et al. 1992, A\&A, 261, 209
 \bibitem[2002]{Her02}
 Herrero, A., Puls, J., \&  Najarro, F. 2002, A\&A, 396, 949
 \bibitem[2004]{Her04}
 Herrero, A., Sim\'on-D\'{\i}az, S., Najarro, F., at al. 2004, proceedings 
 of the International Workshop on Massive Stars in interacting binaries 
 (2004, Quebec, Canada)
 \bibitem[1997]{Hile97}
 Hillenbrand, L.A. 1997, AJ, 113, 1733
 \bibitem[1998]{Hil98}
 Hillier, D.J., \& Miller, D.L. 1998, ApJ, 496, 407
 \bibitem[1997]{How97}
 Howarth, I.D., Siebert, K.W., Hussain, G.A.J., et al. 1997, MNRAS, 284, 265
 \bibitem[1989]{How89}
 Howarth, I.D., \& Prinja, R.K. 1989, ApJS, 69, 527
 \bibitem[1995]{Hub95}
 Hubeny, I., \& Lanz, T. 1995, ApJ, 439, 875
 \bibitem[1978]{Hum78}
 Humphreys, R. 1978, ApJSS, 38, 309
 \bibitem[1980]{Kan80}
 Kane, L., McKeith, C.D., \& Dufton, P.L. 1980, A\&A, 84, 115
 \bibitem[1991]{Kil91}
 Kilian, J., Becker, S.R., Gehren, T., et al. 1991, A\&A, 244, 419
 \bibitem[1994]{Kil94}
 Kilian, J. 1994, A\&A, 282, 867
 \bibitem[1981]{Kuk81}
 Kukarkin, B.V., Kholopov, P.N., Artiukhina, N.M., et al. 1981, CSV
 \bibitem[1979]{Kur79}
 Kurucz, R.L. 1979, ApJS, 40, 1
 \bibitem[1975]{Loh75}
 Lohsen, E. 1975, IBVS, 988, 1
 \bibitem[2000]{Mae00}
 Maeder, A., \& Meynet G. 2000, ARA\&A, 38, 143
 \bibitem[2004]{Mas04}
 Massey, P., Bresolin, F., Kudritzki, R.P., et al. 2004, ApJ, 608, 1001
 \bibitem[2005]{Mas05}
 Massey, P., Puls, J., Pauldrach, A.W.A., et al. 2005, ApJ, accepted, astro-ph/0503464 
 \bibitem[2004]{Mar04}
 Martin, J.C. 2004, AJ, 128, 2474
 \bibitem[2002]{Mar02}
 Martins, F., Schaerer, D., \& Hillier D.J. 2002, A\&A, 382, 999
 \bibitem[2005]{Mar05}
 Martins, F., Schaerer, D., \& Hillier D.J. 2005, A\&A, accepted, astro-ph/0503346
 \bibitem[1962]{McN62}
 McNamara, D.H., \& Larsson H.J. 1962, ApJ, 135, 748
 \bibitem[2003]{Mey03}
 Meynet, G., \& Maeder, A. 2003, A\&A, 404, 975
 \bibitem[2001]{Ode01}
 O'Dell, C.R. 2001, PASP, 113, 290
 \bibitem[1999]{Pal99}
 Palla, F. \& Stahler, S.W. 1999, ApJ, 552, 772
 \bibitem[2001]{Pau01}
 Pauldrach, A.W.A., Hoffmann, T.L., \& Lennon, M. 2001, A\&A, 375, 161
 \bibitem[1985]{Pet85}
 Peters, G.J., \& Polidan, R.S. 1985, IAUS, 111, 417
 \bibitem[1999]{Pre99}
 Preibisch, T., Balega, Y., Hofmann, K.H., et al. 1999, NewA, 4, 531
 \bibitem[2004]{Prz04}
 Przybilla, N., \& Butler, K. 2004, ApJ, 609, 1181
 \bibitem[2005]{Pul05}
 Puls, J., Urbaneja, M.A., Venero, R., et al. 2005, A\&A, in press
 \bibitem[2004]{Rep04}
 Repolust, T., Puls, J., \& Herrero, A. 2004, A\&A, 415, 349
 \bibitem[2004]{Rob04}
 Roberto, M., Beckwith, S.V.W., \& Panagia, N. 2004, AJ, in press
 \bibitem[2002]{Roy02}
 Royer, F., Gerbaldi, M., Faraggiana, et al. 2002, A\&A, 381, 105
 \bibitem[1997]{San97}
 Santolaya-Rey, A.E., Puls, J., \& Herrero A. 1997, A\&A, 323, 488
 \bibitem[1988]{Sch88}
 Schonberner, D., Herrero, A., Becker, S., et al. 1988, A\&A, 197, 209
 \bibitem[2003]{Sim03}
 Sim\'on-D\'{\i}az, S., Herrero, A., \& Esteban, C. 2003, RMxAC 18, 123
 \bibitem[2005]{Sim05}
 Sim\'on-D\'{\i}az, S., \& Herrero, A., in preparation
 \bibitem[1976]{Smi76}
 Smith, M.A., \& Gray, D.F. 1976, PASP, 88, 809
 \bibitem[1992]{Sch92}
 Schaller, G., Schaerer, D., Meynet, G., \& Maeder, A.  1992, A\&AS, 96, 269
\bibitem[1993]{Sta93}
 Stahl, O., Wolf, B., G\"ang, Th., et al. 1993, A\&A, 274, L29
 \bibitem[1996]{Sta96}
 Stahl, O., Kaufer, A., Rivinius, T., et al. 1996, A\&A, 312, 539
 \bibitem[2002]{Tru02}
 Trundle, C., Dufton, P. L., Lennon, D. J., et al. 2002, A\&A, 395, 519
 \bibitem[2005]{Urb05}
 Urbaneja, M. A., Herrero, A., Bresolin, F., et al. 2005, ApJ, in press 
 \bibitem[1996]{Vac96}
 Vacca, W.D., Garmany, C.D., \& Shull, J.M. 1996, ApJ, 460, 914
 \bibitem[2002]{Vil00}
 Villamariz, M.R., \& Herrero, A. 2000, A\&A, 357, 597
 \bibitem[2002]{Vil02}
 Villamariz, M.R., Herrero, A., Becker S. R., et al. 2002, A\&A, 388, 940
\bibitem[1990]{Wal90}
Walborn, N.R., \& Fitzpatrick, E.L. 1990, PASP, 102, 379
 \bibitem[1994]{Wal94}
 Walborn, N.R., \& Nichols, J.S. 1994, ApJ, 425, L29
 \bibitem[1972]{Wal72}
 Walborn, N.R. 1972, AJ, 77, 312


\end{thebibliography}
\end{document}